%% file: PCFrisch.tex
\documentclass[12pt,preprint]{aastex}

\input{PCFdefcommand}

\shorttitle{P. C. Frisch; June 5, 2002}
\begin{document}

\title{The Interstellar Medium of our Galaxy }
\author{Priscilla C. Frisch}
\affil{University of Chicago, Department of Astronomy and Astrophysics, 5460 S.\ Ellis Avenue, Chicago, IL 60637}

\tableofcontents
\newpage

\section{Introductory Comments \label{intro}}

Over the past century our picture of diffuse material in space has
grown from a simple model of isolated clouds in thermal equilibrium
with stellar radiation fields to one of a richly varied composite of
materials with a wide range of physical properties and morphologies.
The solar system interacts with a dynamical interstellar
medium. Optical, radio, and UV astronomy allow us to study the clouds
which form the galactic environment of the Sun. The composition and
distribution of interstellar clouds in the disk and halo tell us about
the history of elemental formation in our galaxy, and the past and
future environment of the solar system.

Dark lanes of dusty clouds obscuring portions of
the Milky Way are celestial landmarks,
but the realization that interstellar gas pervades space is quite recent.
The 20th century opened with the discovery
of a `nebulous mass' of interstellar gas in the sightline towards the
binary star $\delta$ Orionis 
\cite{Hartmann:1904}.  
A series of over 40 spectra showed that the \CaII~K line (3933 \AA)
absorption line was nearly stationary in wavelength, `extraordinarily
weak', and `almost perfectly sharp', in contrast to broader variable
stellar absorption features.  Sharp stationary \NaI~D1 and D2 lines
(5890, 5896 \AA) were discovered in $\delta$ Ori and $\beta$ Sco by
Mary Lea Heger.
An explanation offered was
that a stationary absorbing cloud of vapor was present in space
between these binary systems and the observer.  The \CaII~and
\NaI~lines constituted the primary tracer for interstellar gas during
the first half of the century.

Interstellar matter (excluding dark matter) provides about 
30--40\% of the Galactic mass density in the solar neighborhood.
Trace elements heavier than He, which form the planets, record the
chemical evolution of matter in our Galaxy, and provide detailed
information on physical conditions in interstellar clouds, represent a
small proportion of the interstellar atoms ($\sim$0.15\%).  These same
elements trace the metallicity of interstellar gas, and by inference
the mineralogy of interstellar grains.  A primary goal of interstellar matter
(ISM) studies has been to determine the chemical composition of interstellar clouds
compared to, for instance, normal Population I stars such as the Sun.
Space observations are required to observe most astronomically
interesting elements such as C, N, O, Fe, Mg, Si since the
resonant ground state transitions of these atoms fall in the
ultraviolet (UV, 912--3000 \AA).

Eugene Parker once asked me `What is an interstellar cloud?'  The
Rashomon-like answer depends on the context.  Early optical data
showing velocity components in interstellar absorption lines led to a
definition of 'clouds' as discrete kinematical units.  Alternate
descriptions were based on the physical properties of the clouds,
e. g.  warm diffuse intercloud material in equilibrium
\cite[][]{Stromgren:1948,FGH:1969}, or hot tenuous coronal gas
\cite[][]{Spitzer:1956} to confine the clouds.  Ground and space data
now show interstellar material with densities in the range
10$^{-4}$ atoms \cc~to over 10$^{3}$ atoms \cc, temperatures 20 K
$<$\Tkinetic$<$10$^6$ K, and many levels of ionization.  Within 10 pc
of the Sun, we see density contrasts of over 400 and temperature
contrasts over 100.  The distinction between turbulence and `clouds'
has been blurred by recent high spectral resolution data showing that
low resolution spectral data may miss over half of the velocity
components in a sightline (Section \ref{opticalhires}), and that
$\sim$15\% of the mass of cold clouds is contained in tiny
(\au--sized) structures (Section \ref{patterns}).  Are these features
`clouds', or manifestations of a turbulent ISM?  The discovery of
interstellar clouds in the Galactic halo
\cite{Munch:1957} added new questions about 
the stratification of ISM in the gravitational potential of the Milky
Way Galaxy, and the origin of halo gas.

Surprisingly, interstellar gas constitutes about 98\% of the diffuse
material inside of the heliosphere, and subparsec spatial variations
in interstellar cloud properties near the Sun indicate the solar
environment could change on time scales $\approx$10$^4$ years.
Shapley's conjecture in 1921 that interstellar clouds affect planetary
climates no longer seems outlandish.

Symbols used here are \N\ (column densities, \cmtwo), and \nH\ 
(total volume density for all forms of H, \cc).
Early optical and 21 cm radio data were 
insensitive to clouds with column densities \NH$<$10$^{19}$ \cmtwo.
UV observations of trace elements can detect
kinematical objects with \NH$\gtrsim$10$^{16}$ \cmtwo.
Galactic longitudes and latitudes are quoted in System II.
\footnote{System II was adopted by the International Astronomical Union in
1958 in order to correct earlier errors in the location of the Galactic center.
Galactic coordinates published before 1958 are incorrect.}
The Local Standard of Rest velocity frame represents
heliocentric velocities transformed to the rest frame corresponding to 
the mean motion of comparison stars near the Sun, where the comparison set is
selected according to some criterion.
\footnote{Many of the LSR interstellar velocities presented in the 20th
century assumed, frequently without explanation, a `standard' solar
motion corresponding to a velocity of 19.7 \kms\ towards the apex
position
\glong=57\deeg, \glat=+22\deeg.  Recent Hipparcos results give a solar
motion of 13.4 \kms\ towards the apex direction \glong=28\deeg,
\glat=+32\deeg.  Radio data are usually presented in the LSR velocity frame,
where this issue is particularly troublesome.}

This review focuses on the diffuse gas in the space between the stars of
our Galaxy.  For an eloquent summary of the physical properties of
the ISM and data up to the mid 1970s, see Spitzer's 
book `Physical Processes in the Interstellar Medium' \cite{Spitzer:1978}.

\section{Discovering Interstellar Gas \label{firsthalf}}

The observational and theoretical foundations of ISM space studies
were formed in the first half of the 20th century.  In 1926 Sir Arthur
Eddington laid out the principles of the ionization
equilibrium of atoms in space under the influence of dilute stellar
radiation fields \cite[the `Bakerian Lecture',][]{Eddington:1926}.  He
evaluated the importance of the short wavelength stellar radiation
field ($\lambda$$<$800 \AA) for cloud heating (although extreme
ultraviolet radiation from space was not observable in 1926), and concluded that
diffuse clouds are illuminated by a radiation field at a Planck
temperature of 10,000 K and have kinetic temperatures
\Tkinetic$\sim$10,000 K.  He found frequent collisions would establish
Maxwellian velocity distributions for electrons and ions in space.
Eddington concluded that the material creating the stationary \CaII\
and \NaI\ absorption lines is uniformly distributed, and argued that
stellar dynamics implied $n \leq$10 \cc\ for diffuse material.  He
determined that in space most interstellar Ca is \CaIII\
and most Na is \NaII\ 
He `reluctantly' concluded that dark nebulae derive their obscuration
from `fine solid grains'.  Eddington noted that radiation with
energies greater than $\sim$13.6 eV (the ionization potential of
hydrogen) would be prevented from entering clouds by abundant
hydrogen, and concluded interstellar \HH\ would be abundant.

In the early part of the century, Harlow Shapley advanced the idea
that interstellar clouds were linked to terrestrial climate shifts.
He noted that the diffuse luminous and dark nebulae are found
throughout space, and that the Sun was receding from the Orion region
where dark nebulae are prominent \cite{Shapley:1921}.  Shapley
suggested that a past climate-altering encounter between the Orion
molecular clouds and the solar system would yield a 20\% variation in
solar radiation, which if sustained for a period of time, would alter Earth's
climate.  While encounters with dense clouds as envisioned by Shapley
are statistically improbable,
encounters with clouds of modest density ($\sim$10 \cc) are much more likely and would
destabilize the heliosphere and modify the interplanetary environment
\cite{ZankFrisch:1999}.

\subsection{Optical Absorption Lines \label{optical}}

The interstellar nature of the sharp stationary absorption
features seen in binary systems was quickly established.  
Plaskett and Pearce (1930) provided a convincing discussion that the sharp
\CaII\ and \Na\  lines are formed in diffuse space, and labeled these features 
`interstellar'.  Observations of 1700 stars with $V<$10.5 mag by Otto
Struve had shown that K line strengths in general increase with
increasing magnitude (and thus distance), suggesting an interstellar
origin.  Plaskett and Pearce measured \CaII\ K line velocities for
$\sim$250 OB stars, (to within $\pm$1.8 \kms), and found they are
`almost exactly twice' the interstellar \CaII\ velocities
expected for the galactic rotation of interstellar clouds
located at half the distance of the background star.  Peculiar motions within
the clouds were found to be equally important for line broadening
(Beals 1936).

During the late 1930's, Merrill and collaborators surveyed the
yellow \NaI\ D1, D2 and blue \CaII\ H, K lines in over 400 hot bright stars.
The D2/D1 line ratios were seen to increase as total line intensity
decreased, providing an early indication of line saturation.
The formula relating absorption line equivalent width (\W) and column density
(\N) was found to be faulty for `deep' lines, leading to 
the development of an empirical `curve of growth' (COG) using
the doublet ratio (\CaII\ H/K, \NaI\ D1/D2) to constrain the
functional dependence of \W.
The COG, formalized later by Stromgren (Section \ref{theory}),
has been used to derive column densities from the equivalent
widths of interstellar absorption lines for 70 years.
Cloud motions were shown to depend on star distance, 
with velocity-longitude plots showing a double sine pattern with 
an amplitude at half the expected
Galactic rotation value \cite[e.g., see review of][]{Munch:1968}.
The bulk motions implied by line velocities 
showed a chaotic or turbulent component, with 
dispersion \Vradial=5--10 \kms, which was 
interpreted as moving `clouds' or `currents'.
The \NaI\ line strengths increased with both distance and 
the amount of interstellar `smoke' (now known as dust) in the sightline
(where the dust was determined by the reddening of starlight).

The chemical composition of interstellar clouds was explored by a series of
observations at Mt.\ Wilson, Lick, and the 
Dominion Astrophysical Observatories.  Sharp interstellar absorption lines from
\NaI\ ($\lambda\lambda$3302, 3303, 5896, 5890), 
\CaI\ ($\lambda$4227),  
\FeI\ ($\lambda\lambda$3720, 3860), 
\TiII\ ($\lambda\lambda$3384, 3242, 3229, 3073), 
\KI\ ($\lambda\lambda$7665, 7699),
\CHII\ ($\lambda\lambda$3958, 4233), 
\CHI\ ($\lambda\lambda$4300, 3890), 
\CNI\ ($\lambda\lambda$3876, 3874) were discovered, and upper
limits were placed on 
\AlI\ ($\lambda$3944), 
\SrII\ ($\lambda\lambda$4078, 4216), 
\BaII\ ($\lambda\lambda$4554, 4934) lines
\cite[e.g., see review of][]{Munch:1968}.
The sightlines towards $\chi$2 Ori and 55 Cyg were shown to be
excellent for identifying new interstellar lines \cite{Dunham:1939}.
Dunham used the weak 3302 \AA\ and strong D1, D2 \NaI\ doublets to
construct an empirical COG for the ISM towards $\chi$2 Ori, avoiding
saturation problems with the D lines.  Ionization equilibrium
calculated for \NCaI/\NCaII\ provided the interstellar electron
density, \nel$\sim$0.7--1.4 \cc.
The anomalous properties of interstellar \CaII\ lines were apparent in
early data.  These observations yielded \NCaII/\NNaI$>>$1.5, which was
the value expected from stellar atmosphere data at that time.

The peculiar behavior of interstellar Ca seen in these early data was
a harbinger of the fundamentally different properties of volatiles and
refractory elements in the ISM; however, an incomplete understanding of
cloud ionization prevented this recognition.  When Olin Wilson showed
convincingly that \CaII\ has a velocity distribution which is peculiar
in comparison to \NaI, he described the results as `unexpected...and
disappointing'.  The average internal velocity distribution for \CaII\
was about three times larger than for \NaI\ \cite[22 \kms\ versus 7.5
\kms,][]{Wilson:1939}.  Although later high-resolution spectroscopy
showed that the `lines' observed by Wilson were blends of several
components (Section \ref{opticalhires}), the velocity distribution of
\CaII\ was still peculiar.  Merrill and Wilson showed that the ratio
\RNaCa\ varied strongly from one cloud to another, with values
$<$0.3--10 
\cite[e.g., see review of][]{Munch:1968}.

Walter S. Adams published an influential survey of the strengths and
velocities of \CaII\ H,K, lines and weak features from \CNI, \CHI,
\CHII, \FeI, and \CaI\ lines in $\sim$300 disk (\absZ$<$500 pc) OB
stars visible from the northern hemisphere
\cite{Adams:1949}.
The higher resolution of these data ($\sim$6
\kms), compared to earlier photographic data, demonstrated that
interstellar gas is concentrated in clouds with velocity separations
larger than the atomic velocity dispersion within individual clouds.
The velocities in this survey provided a consistent survey of cloud
kinematics.  Adams estimated the relative strengths of the \CaII\
lines by visually estimating line strengths, which was a common
practice then; the weakest features he detected correspond to
equivalent widths of $\sim$3 m\AA.  Half of the stars in the sample
were found to have complex H or K lines, while about six stars showed
weak \CaII\ absorption at 4226 \AA.  About 25\% of the stars had weak
blue/near-UV features from \ion{CN}{1}, \ion{CH}{1},
\ion{CH}{2}, \ion{Ca}{1}, or \ion{Fe}{1}.
Adams found 21 \CaII\ line components with \Vlsr$>$30 \kms,
\footnote{The velocities quoted in Adams, 1949, 
should be updated with modern wavelengths.}  and concluded the
velocities represented peculiar cloud motions through the local
standard of rest (LSR).

Blaauw used Adams' data to determine the form of the velocity
distribution of the chaotic component of interstellar cloud velocities
\cite[e.g. see review by][]{Spitzer:1968}.  
He compared \CaII\ component velocities with two
possible distributions -- a Gaussian and an exponential distribution
of component velocities.  Blaauw concluded that an exponential form
fit the observed velocity distributions better than a Gaussian form.
Munch extended this analysis to halo stars, with observations of
\CaII\ and/or \NaI\ towards 112 stars in the Northern Hemisphere sky
\cite[l=50\deeg--160\deeg, \absZ=1--2 kpc][]{Munch:1957}.  
He confirmed earlier studies showing that bulk cloud motions indicate
clouds are aligned with the Orion and Perseus spiral arms, and found
blue-shifted absorption components from expanding interstellar gas
around O-star associations.  Based on the Blaauw results
and reasoning that turbulence is not Gaussian so that the increased
energy dissipation by supersonic turbulence would flatten a cloud
velocity distribution in comparison to a Gaussian, Munch fit the
observed disk and halo cloud velocities with the form:
$\Psi$($V$)=$\frac{1}{2\eta}{\rm exp}-|V-V_o|/\eta$ (where $\eta$ is
the mean radial velocity, found to be $\sim$ 5$\pm$1 km/s by Blaauw).
With the exponential distribution, a constant value for the velocity
dispersion for all \NaI\ D2 line strengths is obtained
($\eta\sqrt2\sim$4.6 \kms), providing a better fit to velocity
distributions than a Gaussian form.  This has created what I view as
a conundrum, since the internal velocity distributions of
cold clouds are Gaussian, while the
bulk cloud velocities measured at higher resolution show an
exponential distribution.

A study of the abundance variations between individual clouds was
presented in a classic paper by Routly and Spitzer, in 1952, which
confirmed earlier earlier indications that the distribution of \CaII\
cloud velocities is intrinsically larger than the distribution 
of \NaI\ velocities.  They
found both that $b$(\CaII)$>$1.5\ $b$(\NaI) and that \CaII\ component
velocities are larger than for \NaI.  The systematic decrease of
\RNaCa\ with increasing cloud velocity is known as the `Routly-Spitzer
effect' (RS), manifested as
\RNaCa$\lesssim$1.
RS proposed that the process accelerating \CaII\ also decreased
\RNaCa, either through collisional ionization of \NaI\ during
acceleration, or from relative differences between Ca and Na
depletions onto dust grains in low versus high velocity clouds.  The
first discovery of ISM within 20 pc of the Sun was possible because of
the Routly-Spitzer effect.
\CaII, but not \NaI, was detected towards $\alpha$ Oph (14 pc) \cite{MunchUnsold:1962}.  
Recent data give \RNaCa$\sim$0.1 in the \Vlsr$\sim$--8 \kms\ cloud towards
$\alpha$ Oph \cite[][]{WeltyCa:1996}.  Small scale ( $\sim$1\deeg)
variations in \CaII\ line strengths suggested the presence of 
small clouds ($<$1 pc).  These data provided the first evidence of
small-scale structure and shocked ISM close to the Sun.

As frequently happens, new techniques proven in first-generation
instruments provide limited data, but change the course of
science. The first high spectral resolution (R$>$200,000) observations
of optical absorption lines did not achieve the initial goal of
observing the \NaI\ D line hyperfine components (separated by $\sim$
1.0 \kms), but instead demonstrated cloud velocity structure that is
unresolved in photographic plates.  High-spectral resolution
(R=3--5$\times$10$^5$) spectrometers were developed to observe the
hyperfine \NaI\ lines towards $\alpha$ Cyg using the long focal length
McNath solar telescope
\cite{LivingstonLynds:1964}, and using a Pepsios spectrometer at Lick
Observatory \cite{Hobbs:1965}.  
Much later, after the launch of the \cop\ satellite and after high-resolution
optical data were routinely acquired by Hobbs and others 
(Section \ref{opticalhires}),
the \NaI\ D-line hyperfine splitting was discovered towards $\alpha$ Cygnus,
using a Michelson interferometer 
\cite[effective resolution $R\sim$500,000,][]{WayteBlades:1978}.
Turbulence in cold interstellar clouds was found to be subsonic. 
The +1 \kms\ cloud has
\bdoppler=0.38 \kms\ and a likely temperature of 70--124 K, indicating
turbulent velocities \Vturb=0.26--0.3 \kms\ in comparison to
the isothermal sound speed of $\sim$0.8 \kms.

A thorough and now classic study of the ISM in front of $\zeta$ Oph
(HD 149757) was performed by George Herbig, using the Lick Observatory
120" Coud\'{e} feed spectra with photographic plates \cite{Herbig:1968}.
The star $\zeta$ Oph has a relatively featureless bright continuum
(O9.5Vn, V=2.6 mag, 140 pc, \vsini=380 \kms), and a rich interstellar
spectrum (\ebv$\sim$0.32 mag) lending itself to searches for new
interstellar species.  It is a runaway star from the
Scorpius-Centaurus association moving supersonically through the ISM
(space velocity $\sim$40 \kms), and with a circumstellar \HII\ region
(5\deeg\ radius).  $\zeta$ Oph also has a ram-pressure confined 
stellar wind observed around the star through 60
$\mu$m radiation from heated, swept-up interstellar grains.
Herbig measured absorption lines from
\NaI, \CaII, \KI, \TiII, \FeI, \CHI, \CHII, and \CNI, and set limits
on some 25 additional species at the \W=1--2 m\AA\ level. 
The ISM is contained in two dominant clouds at 
--15 \kms\  and --29 \kms\  (heliocentric velocities).
An empirical curve of growth for the main component (heliocentric velocity, \Vhc=--15 \kms) gave
\bdoppler=2.4 \kms\ and the column densities for these elements.
The photoionization
equilibrium of \NaI, which equilibrates photoionization with the
temperature-dependent recombination rate of Na$^{\rm +}$ and an electron
(\NNaII/\NNaI$\sim$\nNaII/\nNaI,  Section \ref{theory})
gave \nel=0.36--0.54 \cc\ if the --15 \kms\ cloud is $\sim$50 pc from $\zeta$ Oph.
If photoionization of heavy elements (e.g. C) supplies electrons,
densities \nH=500--900 \cc\ result, indicating the cloud is a thin
sheet of thickness $\sim$0.15 pc.  Later studies show Na is depleted
by factors 4--5, raising densities and reducing cloud thickness by the
same factor.  Enhanced abundances of \CaII\ and \TiII\ in the --29
\kms\ cloud identify this material as `intercloud', which Herbig
suggested is local (since it is seen in front of many Scorpius
stars).

\newpage
$~$
\begin{figure}[h]
\vspace*{5in}
\includegraphics{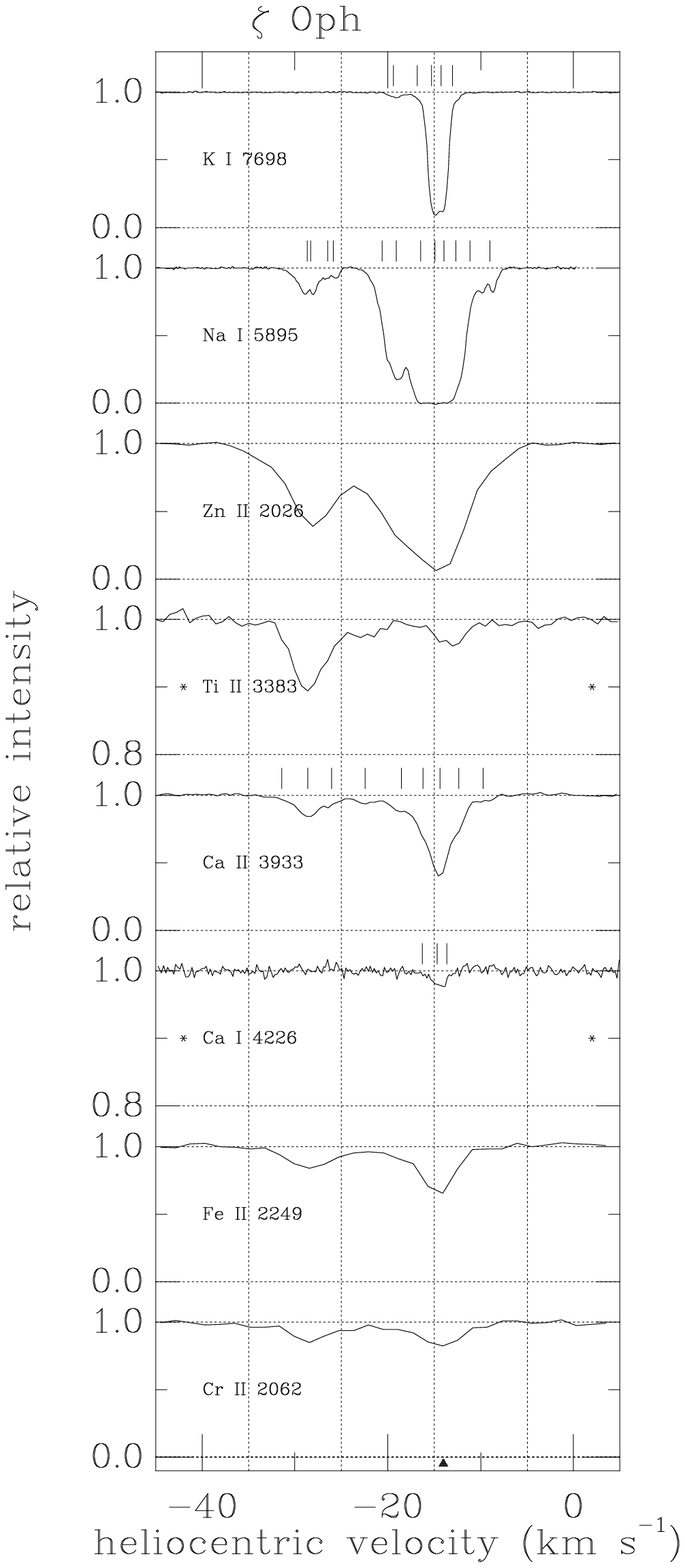}
\vspace*{-0.2in}
\caption{Optical and UV interstellar absorption
lines towards $\zeta$ Oph.  
Cloud A and Cloud B are the prominent absorption features at \Vhc=--26 \kms, --14 \kms, respectively.  
Both \ZnII\ and \TiII\ represent dominant ions.  
The variations in \TiII\ line
strengths between the two clouds are depletion variations, while the
\ZnII\ variations illustrate the differences in the column densities of the
two clouds.
The tickmarks give the central velocities
of components fit to the line.  The triangle on the abscissa represents 0 \kms\ in the LSR.
(Figure courtesy of Dan Welty.) \label{fig-zetaOph}}
\end{figure}

Lewis Hobbs initiated the high-resolution era of optical spectroscopy
with a survey of the \NaI\ D lines at resolution $\sim$1 \kms\
using a Pepsios interferometer at Lick Observatory \cite{Hobbs:1969b}.
Turbulence was found to dominate observed line widths,
since cloud temperatures of 100 K yielded \bdoppler$\sim$0.3 \kms,
versus observed Doppler widths of \bdoppler$\sim$1.5 \kms.
The photoionization balance of Na applied to \NNaI/\NHI\ yielded
typical electron densities \nel$\sim$0.008 \cc\ for solar abundances,
or \nH$\sim$20 \cc\ if electrons are supplied by the photoionization
of metals.  High resolution optical data acquired by Hobbs, his
students, and others proved to be a crucial supplement to UV data (Section
\ref{opticalhires}).

\subsection{Radio Astronomy -- The First Multispectral Data \label{radio}}

Radio waves provided the first multispectral window on the ISM.  
In 1932 Karl Jansky discovered Galactic radio emission 
while searching for radio static sources in a Bell Labs 15 meter
radio antenna.  Jansky recognized the interstellar
origin of the hiss.  Jansky's papers inspired Grote Reber, who, working
with a private radio telescope located in his backyard in Wheaton
Illinois in 1938, detected cosmic static at 2 meters and confirmed
Jansky's discovery.  Over the next several years, Reber used his
backyard telescope to map the northern hemisphere radio sky at 160 MHz
and 480 MHz.  The advent of World War II moved radio astronomy to the
forefront of technical interest, both in the US and Europe, and during
the following decade technical advances from the wartime use of radar
supported the development of the new radio sciences.  
Shklovsky (1960) reviews the scientific basis for the developing field of radio
astronomy.

The spectral index of the cosmic `radio static' was measured and found
to increase towards lower energies, in contrast to the expected
increase towards higher energies predicted for a blackbody (or
'thermal') source.  This puzzle was explained when V. L. Ginzburg
calculated the emission from relativistic electrons ($\sim$1GeV--1
TeV) interacting with interstellar magnetic fields in the halo and
disk, terming the radiation 'magnetobremsstrahlung' 
in analogy to thermal bremsstrahlung from
free-free emission \cite[by free electrons in space plasmas,
e.g.][]{Spitzer:1978}.  Magnetobremsstrahlung is now
generally known as synchrotron radiation.  In 1952 Shklovsky suggested
that a tenuous gaseous envelope must surround the Galaxy in order to
explain the geometry of high-latitude low frequency non-thermal radio
continuum emission.  Baldwin concluded that high-latitude long
wavelength ($\sim$3.7 m) emission originated in a Galactic halo
extending at least $\geq$10 kpc from the plane
\cite{Baldwin:1955}. Interstellar gas in the halo
was considered the carrier for magnetic field which confine halo cosmic rays 
\cite{Shklovsky:1960}.
At low frequencies non-thermal synchrotron emission dominates over 
the optically thick thermal 
emission and the ratio of non-thermal to thermal radio emission 
is \Tbright$\sim$$\nu ^{\rm -1.4}$ (\Tbright\ is the brightness temperature).

During World War II Dutch astronomers deduced the usefulness of radio
emission in probing optically obscured distant portions of the Galaxy,
and predicted that the hyperfine transition of the \HI\ ground \2S12
state would be a useful tracer of \HI\ in interstellar clouds
\cite[e.g.,][]{vandeHulst:1998}.  The hyperfine levels are
statistically populated, $n_{\rm 2}$/$n_{\rm
1}$$\sim$($g_2$/$g_1$)exp(--$E/kT_s$) ($g_2$ and $g_1$ are the
statistical weights of the two levels, $T_s$ is the `spin
temperature').  Although a single spontaneous 21 cm hyperfine
transition is improbable ($\sim$2.85x10$^{\rm -15}$ s$^{\rm -1}$),
this transition is astronomically useful since $\sim$25\% of the mass
density in the solar neighborhood is \HI.  Van de Hulst also
recognized the possibility of stimulated emission in the 21 cm line
through radiative pumping \cite[see, e.g.,][]{Field:1958}.

The discovery of the radio \HI\ 21 cm emission in 1951 at Harvard, in
the Netherlands, and in Sydney, provided a new window on the universe
and made it possible to map interstellar gas and the spiral arm
structure of the Milky Way Galaxy.  Interstellar \HI\ was found to be
strongly concentrated in the Galactic plane and spiral arms, with
emission peaks corresponding to spiral arm velocities.  During the
1950s and 1960s, single-beam surveys of \HI\ 21 cm emission at high
and low Galactic latitudes investigated fundamental cloud properties
for clouds with \logNH$\gtrsim$18.7 \cmtwo.
\HI\ 21 cm emission was used to reconstruct the 
basic spiral structure and kinematical rotation curves of 
our Galaxy 
\cite[e.g., review of][]{Burton:1976}.
The cloud-intercloud nature of the ISM was determined by observations of
21 cm absorption by isolated clouds in front of
bright background radio continuum sources such as the Crab Nebula
and Cassiopeia A \cite{HagenLilleyMcClain:1955}.
In the direction of Cassiopeia A, $\sim$1 kpc away in the Perseus arm,
comparisons of on-source versus off-source data showed three strong
absorbing components corresponding to clouds in the Orion (velocity
dispersion 1.6 \kms) and Perseus (velocity dispersion $\sim$2.5 \kms)
arms.
The first \HI\ shell-like structure in space was discovered in 1958 by
Menon, who observed a shell expanding at $\sim$10 \kms\ in Orion,
centered near the expansion center of the runaway stars AE Aurigae,
$\mu$ Columbae, and 53 Arietis, and with an age consistent with the
formation of these runaways \cite{Menon:1958}.

Cold absorbing clouds (T$\leq$80 K) were found to be common in the
Galactic plane ($|$\glong$|$$<$20\deeg), and broad `intercloud'
emission was seen with no associated absorption, indicating the
presence of both warm and cold gas \cite[e.g.,][]{HoyleEllis:1963,Clark:1965}.  The
median internal velocity dispersion of 2.0 \kms\ is found for cold absorbing
clouds towards 12 sources, versus $T\leq$10,000 K 
for warm emission clouds.  About four absorbing clouds per kiloparsec are found,
while $\sim$7 clouds/kiloparsec are found when interarm regions are excluded.
The high-latitude clouds seen optically at high \Z\ (\Z\ is the distance to the
galactic plane) by Munch and Zirin
were not detected in 21 cm absorption, which was interpreted as
indicating cloud ionization \cite[see review by][]{Munch:1968}.  
Spatial variations in absorption towards
Orion A yielded cloud densities of \nH$\sim$680 \cc\ 

Low column density high-velocity infalling gas (\Vlsr$\sim$--100 \kms)
was discovered at high Galactic latitudes and attributed to the
Galactic halo
\cite[see review by][]{Verschuur:1975}.  
High velocity halo gas
included a \HI\ bridge between the South Galactic Pole and the Large
Magellanic Cloud.  The high velocity clouds (HVC) were recognized as
not-normal disk gas, since at \Vlsr=--100 \kms\ a \Z=500 pc cloud
would fall into the Galactic plane in $\sim$5 Myrs.  Some HVC clouds
were found to be associated with a gas deficiency at lower velocities,
suggesting an origin by the acceleration of low velocity gas
\cite{Verschuur:1975}.

The discovery of pulsars in the late sixties 
provided a powerful diagnostic of the thermal electron component
in the ISM.
The wavelength dependent dispersion of pulsar wave packets is measured 
by the dispersion measure,
\DM$\sim$\dispmeas\ \cc\ pc ($L$ is pathlength), which gives the mean
electron density in the sightline.  The dispersing medium was shown to
be interstellar from the approximate correlation between \DM\ and
pulsar distance over kiloparsec lengths.  An early study of the
latitude dependence of dispersion measures for 36 pulsars concluded
that the thickness of the ionized portion of the Galactic disk
corresponds to \DM$\sim$10--40 pc \cc, yielding
\nelmean=0.03--0.1 \cc, in agreement with values derived from thermal 
absorption of low frequency non-thermal radio background by ionized
disk gas \cite{DavidsonTerzian:1969}.

\subsection{Theoretical Foundations \label{theory}}

Theoretical concepts have played an important role in shaping the
picture of the ISM derived from data.  Stromgren's explanation for
glowing regions of \Halpha\ emission (`Stromgren spheres') surrounding
hot stars located in spiral arms is famous \cite[][]{Stromgren:1939,ODell:1999}.  
Ionized regions which surround hot
stars were modeled using a sharp transition between ionized and
neutral gas, with no leakage of ionizing photons into general
interstellar space.

Competing explanations were presented for the cloud-like nature of the
ISM.  In 1940 Ambarzumian and Gordeladse used the presence of discreet
interstellar clouds defined by dust associated with reflection nebula
to constructed a picture of the ISM with 6--7 clouds per kiloparsec,
each with an average optical depth $\tau$$\sim$0.25.
The role that supersonic collisions between clouds (relative
velocities $\sim$40 \kms) play in creating ionized layers around
clouds and evaporating dust grains was discussed in the 1946 George
Darwin Lecture by Oort.  He concluded that interstellar clouds
must be `continually reshuffled' since grains in both dense and
intercloud regions have the same size, and commented that a large
Reynolds number ($\sim$10$^9$, the ratio of inertial to viscous force)
in the ISM suggested turbulence contributes to the reshuffling
\cite{Oort:1946}. Oort also presented the first, tentative,
discussion of the interaction between an expanding supernova shell
with the surrounding ISM, where he argued that `interstellar friction'
heats and eventually brakes (within $\sim$10$^4$ years) 
shell material.

The interaction of a hot star with a patchy ISM was studied in 1948 by
Stromgren in a basic theoretical paper that also discussed
photoionization equilibrium in a cloudy medium.
Based on a clumpy ISM with clouds filling $\sim$14\% of space
(5--8 clouds per kiloparsec) and the expectation that the space
between interstellar clouds contains material of very low density
(`intercloud' gas), Stromgren concluded photons would leak away from
the immediate vicinity of hot stars, giving a diffuse UV interstellar
field ($\lambda$$>$912 \AA) similar to a T=25,000 K stellar field
diluted by a factor 2$\times$10$^{-17}$.  This field would, in
principle, photoionize low density ISM and cloud rims.  In neutral
clouds electrons are contributed by trace element ionization with
first ionization potential, FIP$<$13.6 eV (primarily C).  For the case
where clouds are far from individual stars (say $>$100 pc), the
general diffuse interstellar radiation field (contributed by many OB
stars) would contain sufficient photons with energies greater than the
Lyman limit (wavelength $\lambda$$<$912 \AA) to ionize rims around
neutral cloud of thickness $\sim$1/\nH$^2$ parsecs.  Stromgren
predicted emission measure \EM$\sim$3 \cmsixpc\ (\EM=\emmeas, where
the integral is over the pathlength \L) for the rims, which was
unobservable at that time, contrasting them to
\EM=9x10$^5$--6000 \cmsixpc\ for O5--B0 star \HII\ regions.  
Stromgren predicted \nH$<$0.08 \cc\ for $\alpha$ Vir using \Na\ data, 
close to the value later found from \cop\ data 
\cite[see references in][]{Jenkins:1987}.  
Diffuse ionized gas is now known to be widespread,
with ionized rims showing \EM=2--10 \cmsixpc\ present on 10\%--30\% of
\HI\ clouds \cite{Reynolds:1995}.
In the 1948 paper Stromgren presented a picture of intercloud material
with density \nH$\sim$0.1 \cc\ filling the space between interstellar
clouds with densities \nH$\sim$10 \cc.  Stromgren also formalized the
`curve of growth' theory, which relates
\N\ and \W\ as a function of cloud temperature which describes the statistical
distribution of atomic velocities.
The key variable is the `Doppler constant' \bdoppler$^2$=$2kT/m$ +
2\Vturb$^2$,
where $T$ is temperature, $k$ is the Boltzmann constant, the atomic
mass is $m$, and \Vturb\ is line broadening due to turbulence.  

The chemical composition of interstellar clouds can be derived from
absorption line data, providing recombination rates and ionization
cross sections are accurately known.  However, atoms may not behave
similarly in interstellar clouds and laboratory vacuum, since
densities differ by $\sim$9 orders of magnitude.  Seaton provided the
first careful evaluation of recombination and photoionization rates
for atoms in space, calculating the ionization equilibrium of \HI,
\CI, \NaI, \KI, \CaI, \HeII, \NaII, \KII, and \CaII\
\cite{Seaton:1951}.  
Seaton concluded that $\chi^2$ Orionis data are consistent with `mean
cosmic abundances', except possibly for calcium which appeared to be
deficient.  For the temperature dependent recombination rate
(typically $\sim$\T$^{-0.7}$), Seaton used Spitzer's temperatures of
\temp$\sim$100 K for \HI, and $T\sim$10,000 K for \HII.  Using the
ratio \NCaI/\NCaII\ from the Dunham observations of $\chi^2$ Orionis
(which is not subject to abundance uncertainties) Seaton determined
\nel=0.066 \cc.  Seaton presented the fundamental principle that both
dust and gas must be included to evaluate the composition of
interstellar clouds, and suggested that metals are concentrated in
dust grains when gas composition departs significantly from cosmic
values.

The empirical identification of interstellar clouds by velocity
components and the sharp outlines of dust clouds against the Milky Way,
suggests clouds are discrete objects.  However, in 1952 Chandrasekhar and
Munch introduced an alternate picture of the ISM in terms of a
continuous distribution of material with statistically distributed
density fluctuations \cite[see article accompanying][S99]{Scalo:1999}.
The microscale of turbulent eddies was characterized by a fundamental
parameter corresponding to the distance where a correlation function
falls to 1/e.  The implications and
influence of this stochastic model of interstellar clouds in contrast
to discreet cloud models is discussed by John Scalo (S99).

The research of Lyman Spitzer, Jr., strongly influenced the study of
the ISM for six decades, and ultimately lead to the launch of UV
observatories into space.  Two oft-cited papers \footnote{Several of
Spitzer's most important papers are reprinted in `Dreams, Stars and
Electrons', along with his personal comments \protect\cite{Spitzer:1997}}
published in the fifties provided proof of the correlation between
abundance and velocity anomalies in \CaII\ (the Routly-Spitzer effect,
Section \ref{optical}) and predicted a hydrostatic highly ionized
Galactic corona.  Spitzer summarized the current state of ISM studies
in the December 1953 Henry Norris Russell lecture of the American
Astronomical Society \cite{Spitzer:1954}.  Spitzer investigated the
putative halo from several approaches.  He made an analogy between
thermally driven winds in the terrestrial atmosphere and warm heated
interstellar gas, and concluded that warm interstellar clouds would be
more buoyant than surrounding unheated material and thus would create
a `Galactic wind' \cite{Spitzer:1954}.  The discovery at Mt. Palomar
of high latitude halo clouds 1--2 kpc away from the Galactic plane,
with velocities $\leq$100 \kms\ \cite{Munch:1957}, raised direct
questions about the source of the confining pressure.  Spitzer made
the bold prediction that highly ionized hot `coronal gas' (\nel$\sim$5
$\times$ 10$^{-4}$ \cc, $T\sim$10$^6$ K) confined these clouds, and
also explained high-latitude radio emission \cite{Spitzer:1956}.
Chandrasekhar and Fermi (1953) had reasoned that the gravitational
pressure of ISM in spiral arms was balanced by the sum of magnetic and
thermal cloud pressures, for field strengths \B$\sim$6 $\mu$G.  Using
a larger thermal pressure, Spitzer concluded thermal and gravitational
pressures balanced, so additional interarm pressure from hot coronal
gas was required to balance magnetic pressures.  The primary tracers
of coronal gas are the abundant highly ionized ions \SiIV, \CIV, \NV,
and \OVI, with UV resonant transitions.  At that time the predicted
emission measure of coronal gas (\EM$\sim$0.1 \cmsixpc) was
undetectable in H$\alpha$ photographic surveys,.  The coronal gas
predicted by Spitzer in 1956 was found by \cop\ $\sim$17 years
later.

The presence of both cold and warm \HI\ raised theoretical questions
about the thermal equilibrium between cold dense clouds and warm
dilute intercloud material.  Thermal stability in a `multiphase
medium' was evaluated in a famous paper by Field, Goldsmith and Habing
\cite{FGH:1969}, using the criterion that heating 
and cooling balance in thermal stability.\footnote{ Many excellent
reviews of interstellar theory are found in the proceedings of a
workshop held in honor of George Field in Elba in 1994
\protect\cite{FerraraMcKeeHeiles:1995}.  The Elba volume presents the
`scientific genealogy' of George Field, listing his advisor, Lyman
Spitzer, Jr.  Carl Heiles was Field's first graduate student. Less
importantly, I was Heiles' first graduate student.}  This requirement
means that for cooling $\sim$ $T^{\alpha}$, $\alpha$ must be $\geq$1,
and cooling increases with temperature.  
The dominant cooling mechanism for cool clouds is temperature-dependent
fine-structure emission from trace elements (such as the \CII\
$\lambda$ 158$\mu$ line).  FGH assumed temperature-independent cosmic
ray heating of clouds.  The result is a thermally unstable phase
between two thermally stable phases on the phase-stability
log(\nH)--log($T$) diagram.  These three phases were humorously
identified as F(ield) for the warm intercloud phase, H(abing) for the
cold neutral phase, and G(oldsmith) for the unstable
phase\footnote{Don Goldsmith now pursues a prominent and stable literary
career sharing the wonders of science with the general public.}.  The
classic FGH paper for modeling a multiphase ISM, along with the
discovery of the Radio Loops, inspired many studies of cloud
equilibrium in a supernova-dominated ISM.  The
FGH two-phase theory has been updated with recent heating and cooling
rates, including photoelectric heating by small grains and large
molecules (PAH's), and cooling by electron recombination on grain
surfaces in warm ($T\geq$8,000 K) gas, and allowing variable input
abundances
\cite{Wolfireetal:1995a}.

\subsection{The Void in Nearby ISM \label{localbubble}}

Our view of nearby interstellar gas is biased by the location of the
Sun with respect to Gould's Belt and the related local void in
interstellar matter now known as the `Local Bubble'
\cite[e.g. see,][]{FrischYork:1986}.  In the 19th century, Benjamin
A. Gould (founder of The Astronomical Journal) realized that nearby
bright stars and young stellar OB associations form a `ring' on the
sky surrounding the Sun (`Gould's Belt').  
Up through $\sim$1980, optical and UV ISM studies were heavily biased
towards clouds associated with the relatively nearby ($<$600 pc)
massive hot stars of Gould's Belt, since
these bright hot rapidly rotating stars provided suitable stellar
continuum for sampling interstellar clouds.
Since hot massive stars tend to be associated with the parent cloud
from which they formed, the ring-like distribution of the stars in
Gould's Belt also defines the ring-like spatial distribution of most
of the mass of interstellar material within $\sim$600 pc of the Sun.
The Sun is located inside of this ring and hence inside a void in the
distribution of interstellar matter in space.  The `Local Bubble'
overlaps the interior of Gould's Belt.

The Gum Nebula borders the Local Bubble.  It is an extended region of
ionized gas originally discovered in \Halpha\ photographs of the
southern Milky Way by Colin Gum in the early fifties
(diameter$>$36\deeg, \dist$\sim$250--500 pc, centered near
\glong$\sim$258\deeg, \glat$\sim$0\deeg, \EM$\sim$1300
\cmsixpc).
The Gum is astronomically rich, with several
stellar associations, over six pulsars, several supernova remnants
(including Vela) and is bordered by the cluster Collinder 121 which
forms one end of the Local Bubble \cite{Heiles:1998}.  The Vela
SNR,$\sim$11,500 years old, is located at a distance of $\sim$250 pc
and is on the near side of the Gum Nebula.

Surveys of the reddening (\ebv) of star light by interstellar dust in
front of nearby B8--A0 stars found that color excesses outline a
vacant region surrounding the Sun, extending $\sim$100 pc in the
directions of the Galactic center and anti-center, and $\sim$50 pc in
the directions of Galactic rotation (\glong$\sim$90\deeg) and
anti-rotation \cite[\glong$\sim$270, see references in ][]{Lucke:1978}.  
An expanding \HI\ ring surrounding the Sun, known as `Lindblad's ring',
has morphology and kinematics suggesting association with
Gould's Belt  \cite[e.g.,][]{Poeppeletal:1981}.
The shell expansion age ($\sim$60 Myrs), elongation, and
morphology are consistent with distortion by Galactic rotation.
Elmegreen argues that both Gould's Belt and Lindblad's ring formed during
the compression of ISM during the passage of the last spiral arm
\cite[about 50 Myrs ago, e.g.,][]{ElmegreenEfremov:1998}.
The intercloud \ebv--\NH\ relationship determined from \cop\ data
(Section \ref{globalcopernicus}), suggests \ebv=0.1 mag, corresponding
to \NH$\sim$5x10$^{20}$ \cmtwo, as the contour boundary approximately
defining Local Bubble `walls'.  A plot of the space-motion of the Sun
through the Local Bubble shows that the Sun has been within this low
interstellar density region for several million years \cite[Figure
\ref{fig-FY83}, and][]{FrischYork:1986}, a fact I do not believe a
coincidence given the sensitivity of heliosphere properties to
interstellar pressure \cite{ZankFrisch:1999}.  By analogy, Sun-like stars with
historically stable astrospheres may be the best
planetary systems for a search for advanced life forms \cite{Frisch:1997,Frisch:1993a}.

\section{The {\it \bf Copernicus}  Era \label{copernicus}}

The dream of flying an ultraviolet spectrometer in space became
politically achievable when cold war politics spawned the space age
and the launch of Sputnik (October 4, 1957).  
Before space UV data on the ISM from sounding rockets and early satellites 
(e.g., OAO-2, ESRO TD-1A), the primary sources of data were optical absorption
lines and radio observations which traced primarily cold and warm \HI.  
The ultraviolet, and later extreme ultraviolet (EUV), energy intervals
provided a window on hitherto invisible ISM.

The \cop\ satellite changed the course of astronomy
with discoveries of deuterium outside of Earth, of the 
non-dominance of highly ionized species in the intercloud medium,  
the association of \HH\ with reddened stars, and variations in the composition
of the gas-phase of the ISM between intercloud and cloudy regions.
These first \cop\ results were published in a series of discovery
papers in the May 1973 Astrophysical Journal. 
Several review articles by members of the original
science team have summarized the scientific achievements
of \cop\ \cite[e.g.][]{SpitzerJenkins:1975,York:1976,Snow:1976}.  
\cop\ found $\approx$6--9 UV velocity components per $\pm$20 \kms\
velocity interval.  The range of ionization, 
temperature and densities discovered by \cop\ for diffuse clouds resulted in the
slow abandonment of the `intercloud medium' concept.
The \cop\ data set remains a unique source of data on
disk clouds, and many of the \cop\ targets have never been reobserved in the UV.

The launch of \cop\ coincided with strong growth in radio and optical astronomy.  
Significant advances
occurred simultaneously in many areas of ISM studies. 
The next satellite with a UV spectrometer was \iue, launched in 1978 as \cop\ was ending its useful scientific lifetime.
\iue\ surpassed expectations by providing extensive data on ISM in the halo
and towards faint disk stars ($V\le$11 mag).  
{\it Hubble Space Telescope} (\hst), launched in 1990,  provided a UV observatory 
with a large collecting area in space.
The spectral resolution of {\it Goddard High Resolution Spectrograph} (\ghrs), 
gave the physical properties of individual disk and halo clouds.
The spectrometers of \iue\ and \hst\ operated over 
a different and broader spectral interval (nominal 1200--3000 \AA) than the \cop\ instrument 
(nominal 912--1400 \AA).  More recently, IMAPS, EUVE, and FUSE provided data 
in the 300$\rightarrow$1200 \AA\ interval.  The understanding we have of ISM in
our galaxy as we emerge into the 21st century represents a synthesis 
of data from diverse sources.

\subsection{Cloudy Sightlines:  $\zeta$ Oph \label{zetaOph}}

{\it Copernicus} conducted unprecedented in-depth studies of the ISM towards $\sim$20 stars,
including moderately reddened stars 
($\zeta$ Oph, $\xi$ Per, $o$ Per, $\zeta$ Per, $\chi$ Oph),
and unreddened or lightly reddened stars 
\cite[$\mu$ Col, HD 28497, HD 50896, $\gamma$ Ara, $\zeta$ Pup, $\alpha$ Vir, $\epsilon$ Ori, $\pi ^5$ Ori, 15 Mon, $\epsilon$ Per, $\delta$ Per, $\mu$ Oph, $\lambda$ Sco, $\beta$ CMa, for references see][]{Jenkins:1987}.
The most influential of these studies was the
first thorough study of interstellar UV absorption lines 
towards the classic ISM target, 
$\zeta$ Oph \cite{Morton:1975}.
$\zeta$ Oph (along with other stars) was also observed by \hst, giving a detailed picture of the
abundances in warm and cold clouds
\cite[][SCS92]{SavageCardelliSofia:1992}.
The $\zeta$ Oph sightline is used as an example, because it is consistently
selected as a target for improving the quality and breadth of ISM studies.
Other stars which sample cloudy regions are discussed in Jenkins (1987).

$Copernicus$ performed a complete scan of the UV spectrum of $\zeta$ Oph, and measured 
$\sim$328 lines of $\sim$40 atoms, over 100 \HH\ lines, and found $\approx$44 
unidentified lines (later attributed to \CI, \HD, or \HH).
The ions \ZnII, \NiII, and \CuII\ were detected for the first time.
The --26 \kms\ and --14 \kms\ cloud blends were later referred to as
`Cloud A' and `Cloud B', respectively (SCS92). 
High resolution optical data ($\sim$0.3 \kms\ FWHM) showed that the 
two main cloud groups towards $\zeta$
Oph represent blends of at least 13 absorption components 
\cite[][]{Barlowetal:1995}.
The 10 components common to both the
\CaII\ and \NaI\ absorption profiles show a range of \RNaCa\ ratios,
corresponding to 
0.5$\rightarrow$2.2 for Cloud A (including the --22.0 \kms\ component),
and 5.6$\rightarrow$400 for Cloud B \cite[][]{Barlowetal:1995}.
Therefore the full velocity component structure 
towards $\zeta$ Oph has not yet been resolved in the UV, and 
the unresolved UV components represent blends of components with different properties.

The chemical abundances of the ISM are found from elemental depletion patterns
(Section \ref{diskISM}).
Morton determined the abundances for over 20 elements and
found sub-solar abundances, by factors of 3$\rightarrow$4,000. 
Gas-phase abundances are generally described in terms of the
`depletion' of the element, \deplX: 
\mbox{$\deplX = \frac{N{\rm (X)}}{N{\rm (H)}}_{\rm obs} / \frac{N {\rm (X)}}{N {\rm (H)}}_{\rm Sun}$.}  The observed and solar abundances for element X
with respect to H (or another undepleted element) are $N$(X)/$N$(H)$_{\rm obs}$ and $N$(X)/$N$(H)$_{\rm Sun}$, respectively.
\hst\ GHRS absorption profiles improved abundance precision considerably. 

The depletion patterns of Cloud A and Cloud B are distinctly different,
and have become a template for defining the properties
of warm disk gas and cold disk gas, respectively
\cite[e.g.][]{SavageSembach:1996}.
The differences between depletions in Clouds A and B are shown 
by the behavior of \TiII\ and \ZnII\ in Figure \ref{fig-zetaOph}.
Figure \ref{fig-zetaOph} shows absorption lines in $\zeta$ Oph of,
from top to bottom, \KI\ (7698 \AA), \NaI\ (5895 \AA), \ZnII\ (2026 \AA), 
\TiII\ (3383 \AA), \CaII\ (3933 \AA), \CaI\ (4226 \AA), \FeII\ (2249 \AA), and
\CrII\ (2062 \AA).  The variations in the \TiII\ line strengths between
Cloud A and B are purely depletion variations, since \TiII\ is the dominant
ionization state.  These variations can be compared with the column density
differences between the two clouds, shown by the HST \ZnII\ line (since
\ZnII\ is undepleted).
Cloud A contains $\sim$4\% of the ISM mass in the sightline, but about
65\% of the \TiII\ atoms (see Section \ref{diskISM}).

In Cloud B, \logNH=21.13 \cmtwo.   The percentage
of H contained in \HH\ ($\sim$62\%) is higher for $\zeta$ Oph than for 
any of the other stars surveyed by \cop. 
\HH\ is shielded from optical pumping in Cloud B, giving
\Tex=56 K from the J=1, 0 levels.  
Cloud B must be sheetlike.  Electron densities calculated from \CI/\CII, \MgI/\MgII, \SI/\SII\ and \CaI/\CaII\ give \nel$\sim$0.7 \cc\
(free from abundance uncertainties). 
A cloud density and thickness of \nH=10$^4$ \cc\ and $L\leq$0.05 pc, respectively,
follow from the assumption that electrons are supplied by the photoionization 
of trace elements (primarily C).
Cloud A includes contributions from the interstellar gas within 14 pc of the
Sun, as is shown by comparative studies of the velocities (Section \ref{opticalhires})
and abundances \cite{Frisch:1981} of $\alpha$ Oph (14 pc) and $\zeta$ Oph.

Typical cold cloud temperatures are known from \cop\ \HH\ data (45$\rightarrow$128 K), \HI\ 21 cm absorption data (typically 25$\rightarrow$75 K), and
the median Doppler broading of \NaI\ and \KI\ ($\sim$80 K). 
Cloud densities in the range of \nH=115--2700 \cc\ and \nel=0.15--0.38 \cc\
are found from \KI\ data, while observations of tiny scale structure in \HI\
21 cm in the ISM give \nH$\lesssim$10$^5$ \cc\ (Section \ref{patterns}
Comparative mean \HI\ and median \HI+2\HH\ column densities for cold clouds are 
\NHI$\sim$10$^{20}$ \cmtwo\ (21 cm data, Section \ref{patterns}), 
\NH$\sim$2 $\times$ 10$^{20}$ \cmtwo\ (\KI, Section \ref{opticalhires}), 
suggesting that perhaps half of the atoms in cold clouds are in
molecular hydrogen.

ISM studies at GHRS resolution fail to distinguished over 60\%
of the component structure (Section \ref{opticalhires}).
These uncertainties are partially overcome by modeling UV data using
velocity information from high-resolution optical absorption lines, 
in a method pioneered by Dan Welty to analyze HST observations of 23 Ori
\cite[][]{Welty23:1999}.
The three components detected by \cop\ towards 23 Ori resolved into 21 velocity 
components in the combined optical/UV data (Figure \ref{fig-23Ori}).
Variations in the spatial distribution between dominant, subordinate,
depleted, and undepleted species prevented the use of a single empirical COG
to analyze line strengths and element distributions.
Subordinate neutral ionization states sample cold dense regions
(\KI, \CaI, \NaI\ in Figure \ref{fig-zetaOph}).
The dominant ionization states of strongly depleted species show the broadest
distribution (e.g. \TiII), reflecting enhanced column densities in warm gas,
and ionization states of mildly depleted species have intermediate \bdoppler\
values.
The fine-structure \CI\ lines give densities
\nH$\sim$10--15 \cc, and a cloud thickness 12--16 pc for the low velocity gas towards 23 Ori.
\cop\ \HH\ data gives $T$=65--150 K, yielding
thermal pressure log(\nH$T$)$\sim$3.1$\pm$0.1 \cc\ K.
The inclusion of turbulence increases pressure by less than a factor of two.

The perils of deriving electron densities from a single species
are illustrated by comparison of \nel\ values derived
from twelve plasma diagnostics in the cold gas gas towards 23 Ori 
\cite{Welty23:1999}.
Photoionization equilibrium calculations for
C, Mg, Al, Si, P, S, Ca, Mn, Fe and Ni, \NaI, and \KI\ yield
calculated electron densities ranging from \nel$\sim$0.04 \cc\ (S, Mn, Fe), to
\nel$\sim$0.95 \cc\ (Ca) and $\sim$0.25 (C, Mg).
The \NCaI/\NCaII\ ratio indicates \nel\ varies by a factor of 2.6 between the two strongest 
cloud components.  
The average properties of the cold low velocity components give
\nel$\sim$0.15$\pm$0.05 \cc, and \nH$\sim$10--15 \cc,
or a fractional ionization of 1\%, indicating that H must be partially
ionized.
The electron density determined from \CIIstar/\CII\ is a factor
of 2--3 less than values derived from \NaI, perhaps indicating
the fine-structure lines are formed in the shielded cloud core.

\begin{figure}[h]
\vspace*{4in}
\includegraphics{PCFfig2.ps}
\caption{Distribution of interstellar clouds towards 23 Ori, illustrating
the distributions of weak trace elements versus strong dominant ions
in high velocity (HV), intermediate velocity (IV), weak low velocity
(WLV), and strong low velocity (SLV) absorption components.
(Figure from Welty et al. 1999) \label{fig-23Ori}}
\end{figure}

\subsection{Intercloud Material \label{lamSco}}

Radio data show that $\sim$60\% of interstellar \HI\ 
atoms are contained in warm clouds ($T >$500 K, Section \ref{patterns}).
Possible ionization sources for the intercloud medium include stellar radiation,
soft X-rays or cosmic rays \cite{DalgarnoMcCray:1972}.
The initial search for intercloud absorption features towards unreddened stars
(\ebv$<$0.03 mag) did not detect highly ionized species.
Observations of bright unreddened stars ($\lambda$ Sco, $\alpha$ Vir,
HD 50896, HD 28497, and $\mu$ Col) 
showed that stars with low mean \HI\ space densities may show foreground
clouds that are small and dense (\nH$\sim$10--10$^3$ \cc), in addition
to low density warm gas.
\cop\ found that \NII\ is widespread in the ISM.
\iue\ and \hst\ provided data
on warm low density clouds towards halo stars, and distant disk stars.

The $\lambda$ Sco sightline (216 pc, \logNHI$\sim$19.5 \cmtwo, \ebv=0.03 mag) 
contains neutral and ionized warm low density interstellar clouds.
$\lambda$ Sco is in the interior of Loop I (Section \ref{patterns}) and samples
nearby interstellar gas towards the upwind direction \cite[see references in][]{Jenkins:1987}.  
Five physically distinct absorbing regions are found.
A warm neutral cloud (\logNH=19.23 \cmtwo, Component 2 shows typical properties for
warm neutral gas, $T\sim$10,000 K, \nH=0.3--9 \cc, and less than 10\% ionization.  
The cloud thickness is $\leq$5 pc.
The spatial distributions of the neutrals \HI, \NI, \OI, \ArI\ are similar due to charge exchange (\OI, \HI) and photoionization (\NI, \ArI).
Component 3 is an extended density bounded \HII\ region (\nHII$\sim$0.1--0.3 \cc)
with diameter $\sim$10--30 pc, \T$\sim$10$^4$ K, and ionization 
consistent with ionization by the general Galactic radiation field with \T$_{\rm eff} \sim$24,000 K (comparable to a B1 star and
the diffuse Galactic field predicted by Stromgren).
The fine-structure populations of \CII\ and \NII\ give \nel=0.1--0.3 \cc,
and \MgI/\MgII\ implies \nel$\sim$0.5 \cc.  
The emission measure of this diffuse \HII\ region towards $\lambda$ Sco 
is comparable to the fully ionized \HII\ filaments detected in \Halpha\ emission
\cite[\emmeas$\sim$1--3 \cmsixpc, e.g., ][]{Reynolds:1995}.
Depletions in the warm neutral and ionized components are similar, log $\delta_{\rm Fe}$$\sim$--1.4 dex.
\MgI\ formation is enhanced in warm, T$\geq$6,000 K, ionized gas where
dielectronic recombination enhances \MgII\ recombination.
Abundances in the \HI\ and \HII\ clouds are similar, including
similar depletions for Fe and Si.  
Component 1 is partially ionized, with similar depletions as the other gas, 
and may represent a ionized cloud $\sim$7 pc thick.  
Components 4 and 5 are highly ionized (with \SIV, and \NSiIII/\NSiII$>$1)
and may represent shocked gas.  

Low column density (\NH$<$10$^{18}$ \cmtwo) intermediate velocity photoionized clouds are a common
component of the ISM. 
For example, \hst\ observations of intermediate velocity gas (IV, \Vlsr=20--50 \kms)
towards disk (e.g. 23 Ori) and halo (e.g. $\mu$ Col) stars show warm (4,000--8,000 K)
tenuous components
of photoionized gas with \NSiIII$>$\NSiII.  Towards 23 Ori, \nel=\nHII=1.5--5.0 \cc,
and the cloud thickness is $\le$0.001 pc \cite[$\sim$200 \au][]{Welty23:1999}.
\cop\ showed that this IV gas covers Orion, indicating filamentary material.

\subsection{Deuterium\label{H2DI}}

The \cop\ discovery that \HH\ is widespread in interstellar clouds was
not a surprise.
However, York's discovery of \DI\ in diffuse interstellar clouds was unexpected.
As a junior member of the \cop\ staff, Don York
was assigned the `uninteresting' stars where no absorption features
had been seen optically.  However, these were the stars where
less saturated lines permitted the most accurate identification of absorption features.
The \DI\ \Lalpha\ line is superimposed on the wings of the \HI\ \Lalpha\ line,
and blueshifted by $\sim$80 \kms\ from the \HI\ \Lalpha\ line center.
In the unreddened star $\beta$ Cen,
the \DI\ line is less saturated, giving the first D/H ratio determined
in interstellar space \NDI/\NHI=1.4$\pm$0.2 $\times$10$^{-5}$ \cite{RogersonYork:1973}.
Hydrogen column densities were determined from
\Lbeta$\rightarrow$\Ldelta\ lines, minimizing uncertainties due to
\Lalpha\ saturation.

Deuterium is produced in Big Bang nucleosynthesis, and
$\sim$50\% of the primordial D is destroyed by astration in stellar interiors.
The early D/H ratio is consistent with ratios derived recently from \hst\
observations of nearby stars (e.g. see papers by Linsky, Vidal-Madjar, Linsky, and 
others).
However, some studies suggest the observed variations in the
D/H ratio between stars are statistically significant and represent real
variations in the chemical composition of nearby ISM 
\cite{VidalMadjarGry:1984}.
The high spectral resolution of IMAPS ($\sim$2.5 \kms) permitted this
question to be revisited \cite{Sonnebornetal:2000}.
Significant variations in D/H are still seen,
from 0.74 $\times$ 10$^{-5}$ in $\delta$ Orion A to
2.2 $\times$ 10$^{-5}$ in $\gamma ^2$ Vel.
These variations, however, do not correlate with 
N/H, as expected for ISM astration in stellar interiors.  Therefore this
very important question is unresolved.  Unresolved saturation in the \DI\
lines may be the culprit in these discrepant results, and this question
may not be settled until an ultra-high resolution 900--1200 \AA\ spectrometer is available in space.

\subsection{Molecular Hydrogen}

Molecular hydrogen probes cold clouds and provides an interstellar thermometer
\cite[see reviews by][]{SpitzerJenkins:1975,ShullBeckwith:1982}.
Theoretical studies predict that most hydrogen is in 
molecular form in reddened clouds.  The formation of \HH\ on grain surfaces
is offset by a two-step dissociation process.
Photons, absorbed into the Lyman and Werner bands, cascade
from the excited electronic states into unbound vibrational levels.
Shielding in cloud cores allows the fraction of hydrogen in \HH, 
\fHH =2$N$(\HH)/[\NHI\ +2\NHH], to approach unity.
For \NHH$>$10$^{19}$ \cmtwo, the \HH\ J=0 and 1 rotational levels typically have strong damping wings which provide unambiguous column densities.
The J=0,1 levels give \Tex\ and \Tex$\approx$\Tkinetic.
Observations of $\sim$61 stars with \NHH$>$10$^{18}$ \cmtwo\ give
a mean cloud temperature of 77$\pm$17 K, and a range of 45$\rightarrow$128 K.

In unreddened sightlines, self-shielding is no longer able to reduce optical pumping.
The detection of complex \HH\ profiles, with apparent broadening of the \HH\
lines with increasing J towards many stars (e.g. $\lambda$ Ori,
$\delta$ Ori A, $\mu$ Col, $\kappa$ Ori, 15 Mon, 30 CMa, $\zeta$ Pup), 
was explained as due to one or more
shortward components that are relatively strong at high J-values and therefore
have high \Tex.   UV pumping models
produce the blueward
components in dense \nH$>$100 \cc, sheetlike ($<$0.02 pc)
clouds that may either be close to the background star
or associated with compressed neutral post-shock gas or thin expanding shells
\cite{SpitzerMorton:1976}.

Molecular hydrogen was found to be widespread in clouds with \ebv$>$0.08 mag, 
where H makes the transition from primarily atomic 
(\fHH$<$0.01) to molecular (\fHH$>$0.01).
For \ebv$<$0.15 mag 90\% of the stars have $\fHH$$<$0.1, and 
for \ebv$\geq$0.15 mag 74\% have $\fHH$$\geq$0.1. 
The mean gas-to-dust ratio for all sightlines is 
\mbox{\NH/\ebv=5.8$\times$10$^{\rm 21}$ atoms} \cmtwo\ mag, 
but stars primarily sampling `intercloud' gas towards disk stars 
(\fHH$<$0.01) yield \\
(\NHI+2\NHH)/\ebv=5.0$\times$10$^{\rm 21}$ atoms \cmtwo\ mag 
\cite{BohlinSavageDrake:1978}.
More recent comparisons between high angular resolution \HI, CO,
and 100 $\mu$m emission show that $\sim$26 infrared cirrus clouds 
show substantial amounts of \HH\ for \NH$>$4$\times$10$^{20}$ \cmtwo\
\cite{ReachKooHeiles:1994}.  Recent FUSE results
show that \HH\ is ubiquitous in the Galactic halo.

The excited J lines were blended at \cop\ resolution.  IMAPS
(with a nominal resolution of $\sim$2.5 \kms) measured
about 70 absorption lines from the Lyman and Werner \HH\ bands in
$\pi$ Sco 
for the rotational levels J=0$\rightarrow$5 \cite[][]{Jenkinsetal:1989}.
However, systematic shifts in \NHH\
reconstructed from strong versus weak lines indicate that 
IMAPS did not resolve all of the absorption components.  
These inconsistent results were resolved if the lines consist of
packed unresolved components with pure thermal broadening at \temp$\sim$80 K.
The high J-level column densities are consistent with population by optical pumping
from radiation originating in nearby hot stars.  

\subsection{The Distribution of Nearby Clouds  \label{globalcopernicus}}

\begin{figure}[h!]
\vspace*{4in}
\includegraphics{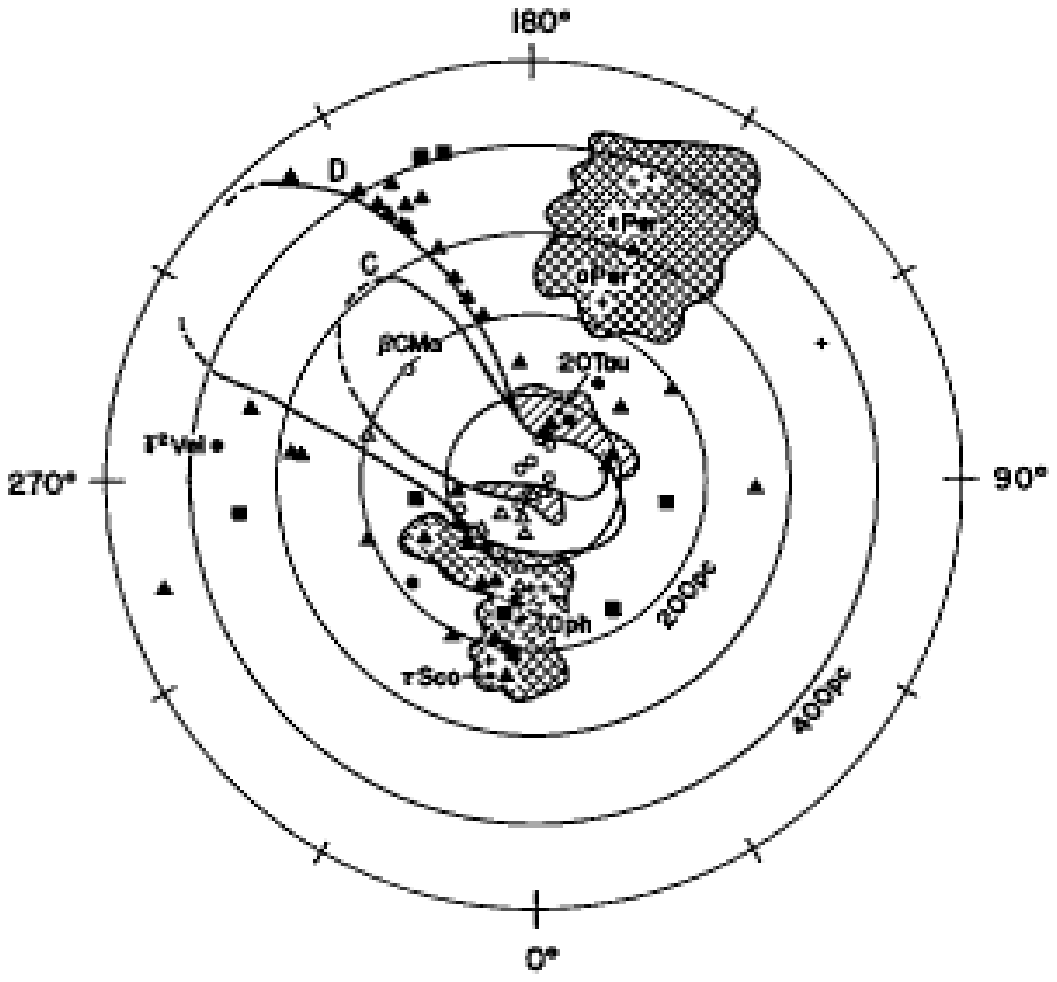}
\caption{Column densities for interstellar hydrogen (\NH) are shown plotted against
star position for stars within 500 pc (from Frisch and York, 1983, with
pre-Hipparcos distances).
The inner contour (C) corresponds to \NH=5 $\times$ 10$^{18}$ \cmtwo,
and the outer contour (D) to \NH=5 $\times$ 10$^{19}$ \cmtwo, where
\NH=\NHI+2\NHH.
The apex solar motion is towards \glong,\glat = 28\deeg,+32\deeg, at
$V_{\rm sun}$=13.4 \kms\
(Section \protect\ref{intro}) indicates that the Sun has been located in region of
space with low interstellar densities for million of years 
\protect\cite[e.g.][]{Frisch:1995}.
Note the void around the Sun corresponding to the Local Bubble.
\label{fig-FY83}}
\end{figure}

Hydrogen is the most abundant element in the ISM, and accurate values 
for \NHI\ could be determined from the 
\Lalpha\ through \Lepsilon\ lines of \HI\ in the \cop\ bandpass.
Accurate \NHI\ values are found from \Lalpha\ data when the damping wings can 
be used, typically for \NHI$>$5 $\times$ 10$^{18}$ \cmtwo)
and stars hotter than $\sim$B1.
A survey of \Lalpha\ and \HH\ 
towards $\sim$100 nearby ($<$1 kpc) stars provided a detailed picture of
the spatial distribution of interstellar hydrogen \cite[e.g.,][]{BohlinSavageDrake:1978}.
The asymmetric spatial distribution of ISM within $\sim$1 kpc in the
disk of the Galaxy, seen originally in
reddening data, and again in OAO-2 data, was demonstrated
by \cop\ data to be a systematic variation of
\nHmean\ as a function of galactic longitude.  The `Local Bubble',
as projected onto the Galactic plane, is shown in Figure \ref{fig-FY83}
\cite[also see][]{Paresce:1984}.

The mean observed H density is \nHmean=$<$\nHI+2\nHH$>$$\sim$1.15 atoms \cc.  
The lowest values are in the third quadrant, \nHmean$<$0.008 \cc.
(The third galactic quadrant corresponds to \glong=180\deeg$\rightarrow$270\deeg, towards $\beta$ CMa.)
Somewhat larger values are found towards Orion, typically \nHmean=0.2--0.5 \cc.
Higher densities are found towards Ophiuchus (e.g. \nHmean=5.5 \cc\ towards $\chi$ Oph), 
and Perseus (e.g.  \nHmean=2.2 \cc\ towards $o$ Per).
Lower densities are found in intercloud sightlines (\fHH$<$0.01),
 \nHmean=0.16 atoms \cc.
Mean densities for \HH\ and \HI\ respectively are
\nHHmean=0.14 molecules \cc\ and
\nHImean$\leq$0.86 atoms \cc\ \cite{BohlinSavageDrake:1978}.
Figure \ref{fig-FY83} shows the strong variations in the mean space 
density of ISM in the nearest 500 pc \cite[from][]{FrischYork:1983}.

\subsection{Chemical Composition of Disk ISM\label{diskISM}}

$Copernicus$ surveyed \HI, \HH, \CI, \SI, \OVI, \MgI, \MgII, \PII, \ClI, \ClII, \MnII, \FeII, \CuII, \NiII, and \ZnII\
across a range of interstellar cloud conditions, and
characterized both the distribution and composition variations of the
ISM within $\sim$500 pc.
Together, \cop\ and optical observations 
measured transitions from two-thirds of the lightest 20 elements
in the periodic table.  The systematic underabundance, compared to solar, of
certain elements in neutral gas could not be attributed to
ionization (but more recently selected solar abundance standards, e.g. O, have
been questioned).  Debates continued on the origin of the
dust grains which are the supposed repositories of the atoms missing
from the gas-phase, as postulated by Seaton.
\iue\ data permitted abundance surveys towards faint disk and halo 
stars, although moderate resolution compromised the abundances
derived from saturated lines.  \iue\ studies of \ZnII\ were important 
as the \ZnII\ 2026 \AA\ line is weak and fell in a sensitive
region of the \iue\ detector.  \hst\ \ghrs\ data provided extensive 
data on abundances in individual warm clouds in the disk and halo, and provided
the detection of (or limits on) additional rare elements (\NX/\NH$<$10$^{-8} $) not 
seen previously (e.g. B, Co, Ga, Ge, As, Se, Rb, Sn, Pb, V, Kr and Te).
The complete picture on elemental abundances in disk clouds provide
data that can be used to test grain formation hypotheses,  
elemental origins, and chemical mixing in the ISM.  
Several excellent reviews discuss ISM abundances \cite[e.g.,][]{Jenkins:1987,Harris:1988,SavageSembach:1996}.

\begin{figure}[h]
\vspace*{4in}
\includegraphics{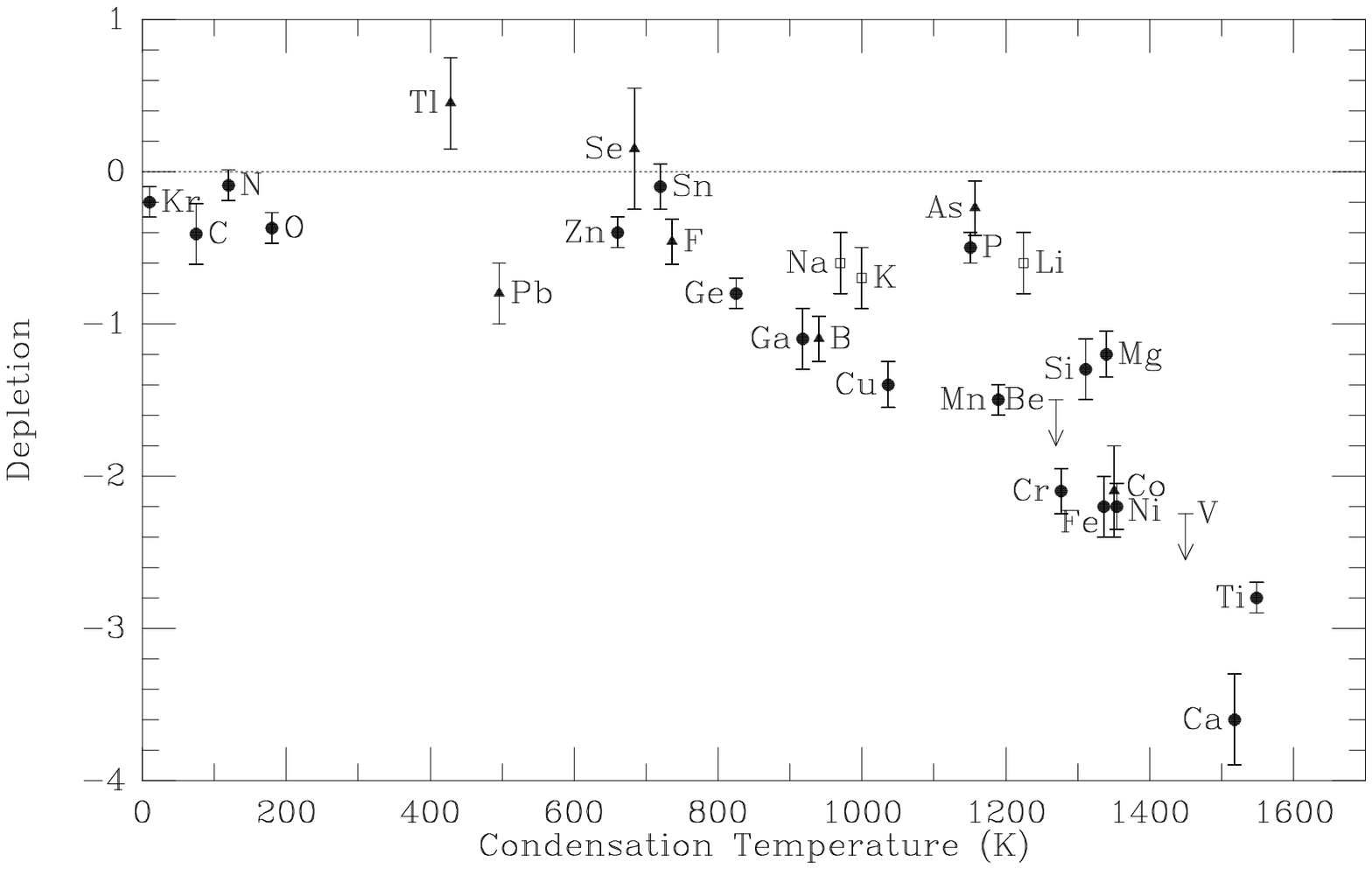}
\caption{Interstellar depletions 
(log $\delta _{\rm X}$=log(X/H)--log(X/H)$_\sun$) in cold disk clouds are shown
plotted against the condensation (\Tcond).  \Tcond\ is the temperature at which 50\% of an
element has been removed from the gas phase through condensation in 
a solar composition a cooling gas under equilibrium conditions.  Variable
abundances for a given value of \Tcond\ (e.g. Mg vrs Fe) reflect the
dominant mineralogy for condensation in stellar environments \cite{Ebel:2000}.
(Figure courtesy of Dan Welty.) \label{fig-depl}}
\end{figure}

Several primary trends in abundance patterns were by discovered by \cop, and confirmed by \iue\ and \hst\ data:  
{\it A}.  The depletion patterns of elements are grouped,
with refractory elements consistently showing greater depletions than
volatile elements.  
{\it B}.  The most depleted elements also show the
largest variations in depletions (which may be a tautology, as Don 
York has pointed out).  
{\it C}.  Refractory element abundances generally anticorrelate with cloud
velocity.
{\it D}.  Volatile elements are generally undepleted in most
sightlines, and may be undepleted in {\it all} sightlines
if component structures are fully resolved and accurate \NH\ values
available.  
{\it E}.  Depletions generally correlate with the mean density
in a sightline, \nHmean, although scatter is present.
These trends are seen in Figure \ref{fig-depl}
({\it A, D}), Figure \ref{fig-deplHWC} ({\it B,D}), Figure \ref{fig-zetaOph} ($C$).
 
Refractory elements (e.g. Fe, Mg, Mn, Ca, Ti, Cr, Ni, Cu, Al, V) 
are characterized by high condensation temperatures and large depletions
(log $\delta$$\leq$1.0 dex in the cold clouds).
In contrast, volatile elements (Kr, S, Cl, Zn, C, N, O, Na, Cl, Ar)
are characterized by low condensation 
temperatures (\Tcond$<$1,000 K) and small depletions.
The volatiles Zn and S are present at solar abundances in
most sightlines (log $\delta <$0.5 dex), and arguments for small
depletions towards reddened stars are unconvincing because of unresolved
component structure.
This behavior of refractory versus volatile elements is evident
in the Routly-Spitzer effect, originally identified (in the 1930's) for optical \CaII\
and \NaI\ lines, although \CaII\ ionization offsets reduced depletions
in warm low density gas.

A \cop\ survey of
\FeII\ towards 55 early-type stars provided a statistically significant sample
which established the anti-correlation between gas-phase abundances of
Fe and \nHmean\ \cite{SavageBohlin:1979}.
Typical values range from log $\delta _{Fe}$$\sim$--1.7 dex towards Orion 
stars (\ebv$\lesssim$0.12 mag) to log $\delta _{Fe} \sim$--2.5 dex towards $\rho$ Oph
(\ebv=0.47 mag).
Ionization is not responsible for these variations, since 
\FeII\ is the dominant ionization state in both warm neutral and warm ionized gas.
Also, warm ionized and neutral clouds with
similar densities and temperatures show similar depletions.
The variable \nHmean\ is also clearly not the best descriptor
of depletion variations, since for example $\zeta$ Oph has two
cloud groups exhibiting dramatically different abundance patterns (Figure \ref{fig-zetaOph}),
but the same \nHmean\ (which is a sightline averaged value).
The large variations in gas-phase abundances found for refractory 
elements is easily understood by noting that $\gtrsim$99\% of the
Fe atoms, for example, are depleted onto dust grains.  The destruction of
$\lesssim$1\% of the dust grains doubles the gas-phase abundance of Fe.

The true test of ISM metallicity will require
accurate abundances for volatile elements in clouds sampling a
wide range of physical conditions.  
B-star abundances are subsolar by a factor of $\sim$2, and may instead 
be the appropriate model for interstellar metallicity \cite{SnowWitt:1996}.
Currently, evidence for solar metallicity in the ISM is limited.
\cop\, \iue\ and \hst\ consistently show that volatiles (Zn, S, Cl, N, O, Kr) 
are present is approximately solar abundances in sightlines with low mean densities.
Studies of Zn ($>$200 stars), Cl ($>$40 stars),
and S ($>$200 stars) show that these elements are depleted by less
than factors of two in low density (\nHmean$<$0.1 \cc) sightlines.
The uniformity of Zn and S abundances at near solar suggest that 
metallicity is relatively invariant within $\sim$500 pc.  
Limited results suggesting that Zn depletion increases with
increasing \fHH\ (e.g. log $\delta_{\rm Zn}$$\sim$--0.5 dex at \fHH=0.1) 
needs deeper investigation of the velocity component structure.
Evidence in favor of subsolar metallicity are observations of Kr,
where observations of weak lines suggest abundances that are $\sim$ 55\% solar
\cite{CardelliMeyer:1997}.
The noble element Kr should not deplete onto dust grains.

Volatile element depletions are
generally independent of mean density for \mbox{\logNHI$<$22 \cmtwo}.
\cop\ observations of \OI\ (1356 \AA) and \NI\ (1160 \AA) towards 53 stars 
found O, N depletions insensitive to \nHmean\ with
ratios \NOI/\NNI$\sim$8 to within 25\%--50\%, and that O and N abundances are 
40\%--70\% solar. 
\hst\ provided C, N, O abundances with systematically lower errors
through observations of weak lines
\cite[see papers by Dave Meyer and collaborators, e.g.][]{MeyerJuraCardelli:1998}.
In clouds with log \NHI=20.2--21.2 \cmtwo,
C, O, N abundances are, respectively, $\sim$39\%, 43\%, 81\% of solar values.
Sulfur is undepleted in warm clouds, but showed small depletions in cold clouds
($\sim$--0.1 dex).
Apparent suprasolar abundances are observed for S in cases where  
\NHI$\sim$\NHII, and ionized gas is not included in estimating \NH.
Chlorine is a special case; it is weakly depleted (factors 2--3), but
\ClII\ reacts rapidly with \HH. 
\NClI/\NClII\ is correlated with \HH\ for clouds where \HH\ is optically thick, while
\ClI\ is unobservable in most other sightlines.

$Copernicus$ studies of the collisionally populated \CI\ fine-structure lines provide an independent diagnostic of cloud pressure
\cite[e.g.,][]{JenkinsShaya:1979}.  
As a trace neutral, \NCI$\sim$\nH$^2$, similar to \NaI\ and \KI\
which correlate tightly with \CI.  However, the correlations
are slightly steeper than linear, possibly indicating that the photoionization rates
may vary between the species, possibly from
shielding of C-ionizing photons by dust in the thicker clouds.
Using the fact that the relative amount of carbon in \CIstar\ versus \CIstarstar\
is a function of the local gas pressure and temperature, Jenkins
and Shaya examined
the parameter range appropriate for neutral clouds.
The carbon fine-structure lines towards $\sim$15 stars yielded
\HI\ gas pressures $<<$10$^4$ \cc\ K, while some show stars showed
pressures  $p/k>$10$^4$ \cc\ K, exceeding the standard
ISM pressure $p/k$$\sim$1,500 \cc\ K \cite{Spitzer:1978}.
Possible sources for high pressure gas include \HII\ regions
near hot stars.

The initial discovery of depletion differences between different elements
led to discussions as to whether this was an effect of the first
ionization potential (FIP) or \Tcond\ of the element.  
Eventually better data resolved this question. Figure \ref{fig-depl}
shows that \Tcond\ is the dominant variable since Zn and Be,
for example, have similar FIP's ($\sim$9.3 eV), yet the depletions
vary by factors of $>$15.
A second puzzle related to the overall variation of depletions with \nHmean.
This correlation was interpreted as support for {\it in situ} grain formation in 
interstellar clouds \cite{Snow:1975}.
Spitzer, however, noted that a statistical sampling of a mixture of warm and cold clouds 
with densities $<$0.2, 0.7 and $>$3 \cc~in the sightline would explain
the observed variations in the depletion-\nHmean~
correlation \cite{Spitzer:1985}.  Spitzer predicted that short pathlengths are more likely to
sample intercloud gas, and intercloud contributions are required to
explain the better correlation between depletions and \nHmean, in
comparison with other parameters such as \NH.

The opposing view presented grain condensation in stellar outflows,
with depletion regulated by \Tcond\ \cite{Field:1974} and reflecting the
grain formation process.  The tight
envelope of the $\delta$-\Tcond\ relation shown in Figure \ref{fig-depl}
supports this later viewpoint.  
However, additional grain modification clearly takes place in the ISM, and mantle accumulation may occur.
In principle, the mineralogy of interstellar dust grains can be derived
from these depletion data.
Ebel (2000) compared 
the condensation sequence of dust grains in a solar composition stellar outflow
with observed depletions (Figure \ref{fig-depl}).
He interpreted the natural groupings of elements with similar \Tcond\ and depletions 
as the result of the formation of
Ca-aluminates  (Ca, Al, Ti), olivine (Si, Mg)
and metals (Fe, Ni, Cr, C,).  The formation of each group corresponds to a phase in the
condensation sequence.  The ability of condensation calculations to explain
the observed depletion groupings, including very different depletions found for 
elements with similar \Tcond\ values (e.g. Mg, Ca), provides strong support
for dust grain formation through condensation in stellar outflows.

\subsection{The Energetic ISM \label{violent}}

Highly ionized gas and high-velocity disk clouds are two aspects of
an ISM component where collisional ionization is significant.
Highly ionized interstellar atoms were expected 
in the ISM, either through production by X-rays or cosmic rays.
The abundances of \SiIV, \CIV, \NV, and \OVI\ peak, respectively, 
at temperatures $\sim$0.6, 1, 2 and 3 $\times$ 10$^5$ K, in collisionally ionized gas.
Stellar evolution supplies energy to the surrounding
ISM through supernova explosions and mass-loss from massive stars.
Routly and Spitzer showed that abundance patterns vary with cloud
velocity, an effect attributed to grain destruction in 
shocks associated with expanding supernova remnants.  Such high-velocity
gas may also be subject to ionization anomalies.

Broad shallow low velocity \OVI\ lines were found to be widespread.
A \cop\ search for lines of \OVI, \CIV, \SiIV, \NV\ towards five stars 
(\dist$\sim$0.1--1 kpc) found that only \OVI\ was generally seen 
\cite{York:1974}.
Combined limits on \NSIV/\NOVI\ and line widths suggest an origin in hot gas with 
2 $\times$ 10$^5$ $<$ \temp\ $<$2 $\times$ 10$^6$ K.
A more extensive survey of \OVI\ towards both reddened and unreddened stars 
found temperature limits \temp$<$3.6x10$^5$ K,
and line broadening and ion ratios consistent with  
formation in hot plasma filling $\sim$20--100\% of the sightline, and
in pressure equilibrium with normal cool clouds 
\cite{Jenkins:1978b}.
About six \OVI\ absorbing regions per kpc were found,
with \NOVI$\sim$10$^{\rm 13}$ \cmtwo\ per region and a dispersion for
bulk \OVI\ cloud velocities of $\sim$26 \kms.  
The velocity distributions of \OVI, \SiIII, and \NII\  are correlated, indicating
the \OVI\ lines is affiliated with less ionized gas 
\cite{Cowieetal:1979}.  
The interstellar nature of \OVI\ was shown by the correlation between \NOVI\
and stellar distances.
Asymmetrical \OVI\ profiles towards  
low halo stars (HD 28497, $\mu$ Col) showed contributions from both local
and halo gas.

The detection of truly interstellar \CIV~and \SiIV\ required
long disk sightlines ($>$1 kpc.
The first detections of \SiIV\ and \CIV\ towards disk stars (\dist=0.3--3 kpc)
found the lines strongest
towards O and Wolf-Rayet stars connected to nebulosity
\cite{Smithetal:1979,Bruhweileretal:1980}.
Although most of the strong sharp \CIV\ and \SiIV\ lines
towards hot disk stars are formed in nebulosity connected with the stars,
broad shallow \CIV\ lines with no associated \SiIV\
appear to arise in the hot interstellar gas producing
\OVI\ \cite{York:1977}.
\hst\ data showed that a diffusely distributed component of
hot gas (\NCIV/\NSiIV$\sim$4.7 and \CIV/\NV$\sim$3.0) also exists in the
Galactic disk \cite{Huangetal:1995}.
Hot gas towards distant disk stars, 
\dist$\lesssim$3 kpc, show a strong correlation between 
\CIV, \SiIV, \NV\ line strengths and star distance. 
The highly ionized and moderately ionized regions are associated,
since the widths of high- and low-ionization lines are correlated.
In the direction of $\zeta$ Oph, three types of highly ionized gas are seen.
Narrow \AlIII\ components originate in
an expanding \HII\ region around $\zeta$ Oph, and broad weak \SiIV\ and
\CIV\ components are detected at --26 \kms\ 
\cite[Cloud A,][]{SavageSembachCardelli:1994}.
GHRS observations of 23 Ori showed that high ions such as \SIII, \SiIV, \CIV, \NV~are associated with both intermediate and low velocity warm and cold clouds
(Figure \ref{fig-23Ori}), suggesting contributions from cloud interface regions.

The frequency of optical high velocity clouds 
and \HI\ 21 cm shells motivated searches for the UV signature of these features.
$Copernicus$ determined that high velocity clouds (\absVlsr$>$100 \kms) 
are widespread in the disk,
including towards active star-forming regions such as Vela and Orion-Eridanus, 
and also towards field disk stars.
Two out of thirteen nearby associations (Orion and Carina)
show high velocity shells with expansion velocities $>$100 \kms\
and ages $\sim$400,000 years \cite{Cowieetal:1981}.
Evidently recombinations lag cooling in these previously shocked clouds. 
\hst\ observations of HVCs towards 23 Ori, Vela, and
$\mu$ Col found supraionized gas, showing ionization levels consistent with 
cloud temperatures of $\sim$25,000 K, but line broadening consistent with 
kinetic temperatures \temp$<$12,000 K \cite[e.g.,][]{Traperoetal:1996}.
The discrepancy is resolved if the HVC represent 
a cooling radiative shock.  The HVC gas is seen in front of most of the
Orion stars, and has been denoted `Orion's Cloak'.
\NCIIstar/\NCII\ yields \nel=\nH=0.4--0.5 \cc~for $T$$\sim$8,000$\pm$2,000 K,
or pressure log($2$\nel$T$)=3.7--4.0  \cc~K.
If carbon is undepleted, the individual HV components have 
thickness 0.005--0.12 pc (1--24$\times 10^3$ \au).
Several superbubbles were investigated by \iue\ and GHRS,
including the nearby Radio Loops I and IV 
\cite[e.g.,][]{SembachSavageTripp:1997}.
The superbubble properties are similar, showing strong \SiIV, \CIV\ and
weak \NV.  Within the bubble interiors, \CIV\ is strongly 
enhanced compared to \NV.
Sightlines sampling tangential directions through the shells (e.g. Loop IV)
show relatively normal high ion ratios, but broader lines.
In many cases, high-ionization gas is kinematically
associated with low-ionization species (e.g. \NI, \SII, \SiII, \FeII) 
which have narrow absorption features at intermediate velocities.

Intermediate velocity clouds (IVC, 20$<$\absVlsr$<$60 \kms) 
are visible in \NI, \NII, \SiIII\, \CaII\ and \NaI\ 
\cite{CowieYork:1978b}.
The ratios \NFeII/\NSII\ and \NSiII/\NSII\ 
correlate positively with increasing
cloud velocity, indicating grain destruction processes 
such as produce the Routly-Spitzer effect \cite{ShullYorkHobbs:1977}.

\subsection{The Nearest ISM and the Local Interstellar Wind  \label{lisw} }

Prior to \cop\, the only observed features in nearby ISM were
\CaII\ lines in a small region towards $\alpha$ Oph (Section \ref{optical}).
\cop\ observations of the nearest stars (1.3--20 pc) showed that low density
(\nHmean$\lesssim$0.1 atoms \cc) interstellar
\HI\ is present in front of all known stars,  
including $\alpha$ Leo \cite[24 pc, \nHmean$\sim$0.02 \cc,][]{RogersonIII:1973}, and $\alpha$ Bootis \cite[11 pc, \nHmean$\sim$0.02--0.1 atoms \cc, ][]{Moosetal:1974}.  
Contemporary with \cop, Boksenberg et al. used a balloon-mounted
spectrograph to observe \MgI\ and \MgII\ in the 
direction of $\alpha$ Leo, yielding the first estimate of the
electron density in nearby interstellar gas 
\cite[\nel$\leq$0.6 \cc\ at $T$=10$^4$ K, ][]{Boksenbergetal:1975}.
The nearby ISM was observed in nearby 
stars of A, B spectral types \cite[e.g. $\alpha$ CMa, $\alpha$ Lyr, $\alpha$ Gru, $\alpha$ Leo][]{RogersonIII:1973,Kondoetal:1978},
and cool (G, K) stars where interstellar \Lalpha\ absorption is
superimposed on chromospheric emission features 
\cite[$\epsilon$ Eri, $\epsilon$ Ind, $\alpha$ Aur, and $\alpha$ Cen A, e.g.][]{McClintocketal:1978,Landsmanetal:1984}.
Interstellar lines superimposed
on chromospheric emission require the uncertain analysis step of modeling
unattenuated stellar \Lalpha\ emission based on observations of solar \Lalpha,
which is the only unattenuated stellar \Lalpha\ that can be observed because
of the strength of the interstellar \Lalpha\ feature.
Nevertheless, a consistent picture emerged showing that the Sun is embedded in a low density 
(\nHI$\sim$0.1--0.15 \cc) warm ($T\sim$7,400 K) interstellar
cloud extending to at least $\sim$3.5 pc from the Sun in all directions.
Observations of the unreddened stars $\alpha$ Vir and $\lambda$ Sco 
found \T$\sim$7,000 K for clouds that are 1--10 pc in size and probably
embedded in the LISM.
Nearby ($<$30 pc) interstellar gas has an asymmetrical distribution
\cite[Figure \ref{fig-FY83}, and][]{Genovaetal:1990}. 
This asymmetry mimics the asymmetry of the Local Bubble
in the sense that the lowest column densities through the Local Fluff
are towards the third Galactic quadrant and North Galactic Pole, and this asymmetry
results from a flow of gas away from Loop I \cite{Frisch:1981}.

On a personal note, I found it intriguing
that the velocities of interstellar \HI\ and \HeI\ inside of the solar system 
are close to the \CaII\ velocity towards the nearby star $\alpha$ Oph, 
and thought that the Sun might inside be inside of the $\alpha$ Oph  cloud.
As a result, I
proposed observe the interplanetary \Lalpha\ glow with \cop.
\footnote{However, I did not dare include this motivation in my
observing time proposal, since at that time ISM was what 
was viewed towards distant stars, and not part of the solar system.
Don York, who organized the \cop\ Guest Investigator program, told me that
before approving my proposal
he had to do some library research to verify that
interstellar gas really is observed inside of the solar system.}
The results were the first spectral observations of the \Lalpha\ glow, 
which established the relation between ISM inside and outside of the solar system \cite[][]{AdamsFrisch:1977}.
Thus \cop\ obtained the first spectrum of interstellar \Lalpha\ emission 
within the solar system during the 1975 solar minimum conditions.
We found a velocity for the `local interstellar wind' in good agreement with interstellar velocities towards
several nearby unreddened stars (e.g. $\alpha$ Vir).

$Copernicus$ also made the surprising discovery
that the \HI\ \Lalpha\ line towards the nearest star $\alpha$ Cen (1.3 pc)
is redshifted by $\sim$8$\pm$2 \kms\ with 
respect to the unsaturated \DI\ line in the same star \cite[e.g.][]{Landsmanetal:1984}.
This shift, since confirmed by IUE and HST, was originally interpreted as flagging the existence of two
clouds in front of $\alpha$ Cen with the Sun located
near the boundary between the clouds.
The heliosheath contribution to this line was recognized later
\cite[e.g.][]{LinskyWood:1996,Gayleyetal:1997}.

The nearest ISM is nicknamed the `Local Fluff', 
a term first used by Don Cox 
at the Local Interstellar Matter 
COSPAR meeting held in Toulouse France in 1986. He was 
trying to describe the tenuous nature of the interstellar gas surrounding 
the Sun, and finally waved his hands and said `it's just sort of this 
local fluff'.  The `Local Fluff' represents the ensemble of interstellar 
clouds (or cloudlets) within $\sim$30 pc of the Sun, regardless of cloud velocity.  (The proceedings of this meeting, and a related heliosphere session, offer an early look at
research into the properties of the LISM.)

The ISM nearest to the Sun is our best sample of warm partially ionized
diffuse gas. 
HST studies of the nearest ISM show typical mean densities $<$\nHI$>  \sim$0.1 \cc\ and temperatures $<T>  \sim$ 7,000 K.
Observations of \MgII/\MgI\ and \CII\ fine-structure lines yield
ionization levels of \nel$\sim$0.12 \cc, which are well modeled by photoionization models
which include EUV radiation from an interface between the Local Fluff gas and
hot plasma.
The LISM gas shows the abundance pattern of shocked gas, with
$\delta_{Fe}  \sim-0.67$ dex$\rightarrow$--1.34 dex locally \cite{Frischetal:1999}.
A recent theoretical model of radiative transfer in local gas shows 
cloud properties corresponding to T$\sim$7,000 K, \nHI$\sim$0.24 \cc\ and
\nel$\sim$0.13 \cc\ \cite{SlavinFrisch:2002}.

There are several outstanding summary publications discussing nearby ISM 
\cite{KondoBruhweilerSavage:1984,CoxReynolds:1987,Frisch:1995,Ferlet:1999}.

\section{Ground-Based Astronomy during the Space Age\label{ground}}

The launch of the \cop\ satellite coincided in time with
several important advances in ground-based astronomy,
including the discoveries of giant shells of radio continuum emission,
of pervasive components of tenuous neutral gas, diffuse plasma, 
tiny `cloudlets', 
and of a highly structured neutral component 
seen in absorption in both optical and radio data. 
Ground-based astronomy benefited from large telescopes, long exposure times and flexibility in detector design, providing an important synergy with UV data.

\subsection{The Radio Sky \label{patterns}}

Radio astronomy supplies two-dimensional morphological data on the
distribution of neutral and ionized gas in space, and provides a
key perspective on the structural characteristics of the ISM,
including determining the spiral structure of our Galaxy \cite[see review by][]{Burton:1988}.

The inhomogeneous \HI\ 21 cm sky
led to the search for the lowest column density sightlines out of the Galaxy.
In the Galactic plane, the direction showing minimum column density 
(\NHI$\sim$4.5 $\times$ 10$^{21}$ \cmtwo) is towards the sightline 
\glong=245\deeg$\pm$6\deeg and \glat=3\deeg$\pm$6\deeg\
\cite{StacyJackson:1982}.  
This minimum, at the edge of the Gum Nebula and known as the `Puppis Window',
 occurs as a $\sim$ 6\deeg\ hole in the local gas 
 at \Vlsr=--10 to +5 \kms.  The Puppis Window samples an elongated direction through 
the third quadrant void (seen in reddening data, Section \ref{localbubble}),
and contains the stars $\beta$ CMa and $\epsilon$ CMa, known for their exceptionally 
low column densities \cite[][]{FrischYork:1983,Paresce:1984,WelshTi:1997,Heiles:1998}.
The Puppis Window appears to result from overlapping shells,
including the 0.8 kpc distant supershell GSH 238+00+09.
(Another name for the Puppis Window is the `$\beta$ CMa tunnel'.)

The Rosetta Stone for understanding \HI~shells, anomalous
\CaII~abundances, the preponderance of negative velocity
\CaII~components, and the Routly-Spitzer effect was
the discovery of large loops of radio continuum emission, especially Loop I
encompassing the North Polar Spur.
Four nearby ($<$250 pc), high-latitude, non-thermal, radio continuum shells 
(408 MHz) were found with surface brightness consistent with 
the brightness-diameter relation found for supernova remnants 
\cite[SNRs, e.g.][]{BerkhuijsenHaslam:1971}.
The Radio Loops are formed by synchrotron emission from cosmic
ray electrons interacting with the interstellar magnetic fields
that have been compressed into giant shells by expanding supernova remnants.
In a widely influential talk at an IAU symposium in 1979,
Harold Weaver linked the Loop I radio continuum shell with filamentary \HI\ seen
in 21 cm emission, and concluded both were formed by a large supershell
around the Scorpius-Centaurus Association. 
The large nearby Radio Loop I (distance$\sim$130 pc, radius $\sim$116 pc),
also known as the North Polar Spur (NPS),
dominates ISM in the northern hemisphere sky and is associated with a slowly
expanding \HI\ shell ($\sim$20 \kms) and a bright source of X-ray emission.
Loop I is a supernova remnant in the radiative phase.
The radio continuum, \HI~and X-ray emission from Loop I
are spatially separated, with ionization decreasing from the interior
to exterior of the shell.
Comparison of rotation and emission measures, Zeeman splitting,
and \HI\ 21 cm data
yield best values of \nel$<$0.4 \cc\, \Bparallel$\sim$1.2--6 \muG,
 \nHI=4 \cc, and $T\sim$100 K (21 cm data) in the 
NPS \cite[e.g.][]{Heiles:1989}.  
Magnetic pressure $B^2/8\pi \sim$4,000 K \cc\ dominates the thermal pressure
of neutrals, and indicates an unstable gas.
This latter point is important because generally magnetic pressure is neglected.

The direction with minimum \NHI~towards the Galactic poles
is towards \glong=150\deeg, \glat=$+$53\deeg, where \logNHI=19.65 \cmtwo\
\cite[the `Lockman Hole',][]{DickeyLockman:1990}.
Four \HI~velocity components are seen towards this minimum 
(0, --10, --50, and --100 \kms),
with most \HI~mass at $\sim$ 0 \kms.
The stratification of \HI\ perpendicular to the Galactic plane in this
direction indicates that 
halo gas (\Z$>$500 pc) contains $\sim$13\% of the \HI\ and corotates with disk gas.  
The \HI~layering can be described as the sum of three distributions.
The CNM distribution can be described by a Gaussian with a scale height dispersion of $\sim$100 pc.
The WNM can be described by the sum of two distributions:  a Gaussian with scale height dispersion $\sim$250 pc, and
an exponential with scale height $\sim$500 pc 
\cite[][Table \ref{tab-scaleH}]{DickeyLockman:1990}.

Observations of \HI\ in absorption versus emission have
provided an increasingly precise picture of the
relative distributions of cold and warm \HI\ gas. 
The first large-scale surveys of 21 cm absorption towards
bright radio sources yielded a statistical picture of the components
of the two-phase ISM
\cite[e.g.][]{HughesThompsonColvin:1971,RadhakrishnanGoss:1972}. 
The equivalent thickness of the Galactic plane for cold absorbing clouds
was 330 pc,
with a midplane density of 0.29 \cc\ for an assumed Gaussian distribution.
The WNM  extended to larger scale heights, with thickness 585$\pm$100 pc, and lower midplane densities (0.155 \cc).
Cold (60--80 K) absorbing clouds were found to contain
35--45\% of the \HI, and appear to be confined mainly to spiral arms.   
The \HI~absorbing clouds have mean column density
\NHI/\Tspin=1.5 $\times$ 10$^{19}$ atoms \cmtwo~K$^{-1}$ kpc$^{-1}$ and
a harmonic mean temperature $T$=71\deeg$\pm$9\deeg.  
Typically over 2.5 cold clouds per kiloparsec are found for
a sightline along the Galactic plane.
The emission is contained in broad features formed in ubiquitous warm intercloud \HI.
Mean spatial densities for the cold and warm clouds, respectively, are
\nHmean$\sim$0.7 \cc\ and \nHmean$\sim$0.25 \cc.
The temperatures found for the WNM are uncertain, since
a single unrecognized cold ($\sim$80 K) cloud in the sightline containing
one-tenth of the warm cloud column density would
decrease the WNM temperature derived \Tspin~to 800 K \cite{KulkarniHeiles:1987}.

With the discovery of highly-ionized gas, the concept of a two-phase medium fell out of favor, 
and the term `not strongly absorbing' gas (NSA) was introduced
to distinguish the kinematically broad \HI~21 cm emission components ($T\sim$5,000 K)
from narrow cold absorbing components \cite{DickeyTerzianSalpeter:1979}.
The NSA gas was attributed to either
warm surface layers on cold clouds, or independent clouds.
NSA clouds have an asymmetric velocity distribution,
with --20 \kms~and +10 \kms~components equally frequent,
possibly flagging infalling gas since many NSA sightlines are towards the North Galactic Pole region.
The exponential form of the high-velocity tail was similar to 
the distribution of \CaII\ components seen optically (Section \ref{optical}).
NSA gas is deficient in the interior of the Local Bubble surrounding the Sun, 
and beyond that forms $\sim$ 35\% of the observed \HI\ \cite{Heiles:1980}.
Large holes in the NSA \HI\ are common (diameters$\leq$400 pc), evidently 
corresponding to the interiors of shells surrounding stellar
associations.  Heiles showed the absence of NSA is associated with
anomalous velocities and anomalous velocity dispersions for \HI, and
may represent ISM swept up by expanding shells of gas
which have recombined before deceleration.

Recently Heiles (2001) has used high signal-to-noise data on Zeeman splitting
of the \HI\ absorption line to produce accurate temperatures for
cold clouds.  He found that 54\% of the CNM clouds (containing 61\% of
the total CNM column density) have 
temperatures in the range of \temp=25$\rightarrow$75 K.
Colder clouds, \temp=10$\rightarrow$25 K, are also present 
and contain $\sim$11\% of the mass.
About 40\% of the WNM components have \T=500$\rightarrow$5,000 K, containing
$>$47\% of the total WNM column density.  
About 60\% of the \HI\ is WNM with \temp$>$500 K.
The mean column densities for WNM and CNM, respectively, 
are $\sim$2.2 $\times$ 10$^{20}$ \cmtwo and 
$\sim$0.8  $\times$ 10$^{20}$ \cmtwo.
The WNM temperature range (\T=500$\rightarrow$5,000 K)
coincides with the thermally unstable range which separates CNM from
WNM in the MO theory.

The discovery of radio continuum loops, \HI\ filaments, and \HI\ shells
led to a new class of ISM models, based on the injection of energy into the 
ISM from expanding supernova remnants.
Models included overlapping remnants which create tunnels of hot gas in space \cite{CoxSmith:1974},
Galactic fountains which release hot buoyant gas into the low halo
\cite{ShapiroField:1976}, 
models of cold clouds evaporating inside of a supernova remnants \cite[][MO]{McKeeOstriker:1977,McCraySnow:1979},
and models of expanding superbubble shells sweeping up the ISM \cite{MacLowMcCray:1988}.
The MO multiphase model predicted successive layers of hotter and more ionized
gas on the surfaces of cool clouds.  In this equilibrium picture, 
SNR evolution is altered by the evaporation of embedded clouds, and a
large fraction of the disk is filled with hot ionized material
(HIM, \T=4.5$\times$10$^5$ K, n=0.0035 \cc, $\chi$=\nel/\nH=1.0).
MO predicted the outer regions of standard clouds of cold neutral material
(CNM, \T=80 K, n=42 \cc, $\chi$=10$^{-3}$) are ionized by diffuse stellar
UV and soft X-ray photons, producing layers of warm ionized material
(WIM, \T=8,000 K, n=0.37 \cc, $\chi$=0.15),
and warm neutral material (WNM, \T=8,000 K, n=0.25 \cc, $\chi$=0.68)
near cloud surfaces.  Models of the ISM are summarized nicely
in articles by Cox and McKee in the Elba volume \cite{FerraraMcKeeHeiles:1995}.

A seminal advance in understanding high-latitude ISM resulted from the
combination of Northern and Southern Hemisphere 21 cm emission data, to create a series
of `maps' of the 21 cm sky ($\sim$0.5\deeg~resolution) at discreet 
velocity intervals for \absglat$>$10\deeg\ \cite[e.g.,][]{ColombPoppelHeiles:1980}.  
The recent standard for a 21 cm survey is the 1 \kms\ resolution Leiden-Dwingeloo survey
of northern hemisphere \HI\ between --450 \kms\ and +400 \kms\ \cite{Hartmann:1997}.
The ISM in the solar neighborhood
exhibits rich structural complexity, with most material contained giant
loops, arcs, `worms', and incomplete shell-like features offering little
support for the classic  Ambarzumian picture of an ISM 
filled with spherical clouds. 
Widespread topless shells, capable of supplying hot gas to the halo, are seen
throughout the galaxy.
The nearest example of a worm is the North Polar Spur.
These maps, and subsequent 21 cm studies by Heiles and collaborators, showed
that the ISM in the solar neighborhood is superbubble dominated.
Over 100 expanding \HI~shells and superbubbles are detected, with radii$<$1.2 kpc,
masses 2$\times$10$^7$ \Msun, and expansion velocities $<$24 \kms\ 
corresponding to energies $\geq$10$^{53}$ ergs \cite[e.g.][]{Heiles:1982}.
\HI\ shells are clearly associated with star formation regions (Orion, Carina, Per OB2, Cep OB3, Sco OB3, Loop I),
although only $\sim$33\% of the radio-emitting supernova
remnants also exhibit \HI~shells.
\HI\ gas in shells is significantly cooler ($T$=35--200 K and \nH$\sim$2 \cc)
than outside of shells, and the
expanding shells associated with Radio Loop I and Eridanus are in statistical equilibrium since shell ages ($\sim$2 Myrs) exceed cooling times
for $\sim$2 \cc\ shell gas ($\sim$0.4 Myrs).
These SNRs are also a source of high-velocity gas;
$\approx$15\% of 100 observed radio continuum SNRs (1.5--9 kpc distant) show
high velocity (70--160 \kms) \HI,  probably accelerated by SNR blast waves \cite{KooHeiles:1991}.

The pervasiveness of \HI\ 21 cm high-latitude high velocity clouds (HVCs) 
raised questions about the origin of this gas, and whether the gas is
infalling onto the plane, or ejected from disk supernova remnants. 
Additionally, warped spiral arms in the outer galaxy are seen at high velocities,
leading to the definition that generally gas must deviate
by $>$50 \kms~from predicted Galactic rotation velocities to be considered a HVC.
Several excellent reviews discuss HVCs \cite[e.g.][]{Verschuur:1975,York:1982,WakkervanWoerden:1997,Savage:1995}.
More recently, an extragalactic origin for HVCs from gas falling 
onto the Local Group has been proposed \cite{Blitzetal:1999}.
High and intermediate velocity clouds are grouped into complexes 
tens of degrees across on the sky, with predominantly negative velocities at 
high latitudes.
Up to 37\% of the sky is covered with \HI~gas with \absVlsr$>$100 \kms.
Up to 18\% of the sky is covered with HVC's with 
\absVlsr$>$100 \kms~and \logNHI$>$18.30 \cmtwo.
Low column density HVCs (17.5--18.4 \cmtwo)
tend to surround higher column density HVCs ($>$18.4 \cmtwo), and 
occupy about 19\% of the sky.
Up to 10\% of the mass of \HI~in the 
Galaxy may be in HVC's, and several HVC's have bulk energy comparable to 
that of supernova.  
\cite[e.g.][]{Verschuur:1975,York:1982,WakkervanWoerden:1997,Savage:1995}.

A new class of tiny dense \HI\ clouds were discovered at Berkeley
using VLBI techniques \cite{Dieteretal:1976}, but acceptance of this discovery
was delayed for years because it challenged the two-component model of the ISM.
The search for \HI\ clumps was motivated by
the discovery of tiny ($\sim$10$^{-2}$ \au) regions of ionized gas
inferred from pulsar scintillation data 
\cite[e.g.][]{RickettLang:1973,Cordesetal:1988}.
Observations of 3C 147 showed absorption variations on scales of 0.1".
This discovery was later confirmed using VLBI techniques
\cite{Diamondetal:1989},
and measurements of time-variable 21 cm absorption line in six 
transversely moving pulsars \cite{Frailetal:1994}.
Over 20\% of cold \HI\ is located 
in small dense clumps with scale sizes 5--100 \au\ and 
densities \nHI$\sim$10$^5$ \cc.
The `tiny-scale atomic structures' (TSAS)
are a ubiquitous feature of ordinary cold clouds, and 
are consistent with 
a composite picture of cold TSAS clouds ($\sim$15 K, or
\bdoppler$_{\rm H}$$\sim$0.5 \kms\ in the absence of turbulence)
interspersed with warmer and less dense `inter-TSAS' gas.
The TSAS structures 
occupy only a few percent of the CNM cloud volume, but contribute
10\% of the column density if they have a sheet-like
configuration, or 30\% if they have a filamentary configuration
\cite{Heiles:1997}.
Absorption lines have been seen to vary in strength between 
both members of a binary system, 
and towards the same star observed at multiyear intervals 
\cite[e.g.][]{LauroeschMeyerBlades:2000}.  These variations imply structure 
in the ISM on scale sizes of 10$^1$$\rightarrow$10$^{3}$ \au.

\subsection{High-Resolution Optical Data\label{opticalhires}}

The transition from photographic plates to photon-counting
spectrometers (based on Digicon and CCD technology)
occurred during the 1970's, allowing high-resolution ($\leq$0.5 \kms), 
high signal-to-noise ($W<$1 \mAA) absorption line studies.  
Although only a few species are observed optically, ultra-high spectral
resolution optical data provide a useful template for the component structure 
and depletion patterns in lower resolution UV data.
Early surveys by Hobbs and his students provided the first information
about the distributions of \CaII, \CaI, \TiII, \NaI, \KI, and \LiI, and the
physical properties of the gas giving rise to these lines.
Recent surveys benefit from ultrahigh spectral resolutions, high
signal-to-noise, and good quantitative analysis tools.

\begin{figure}[h]
\plotone{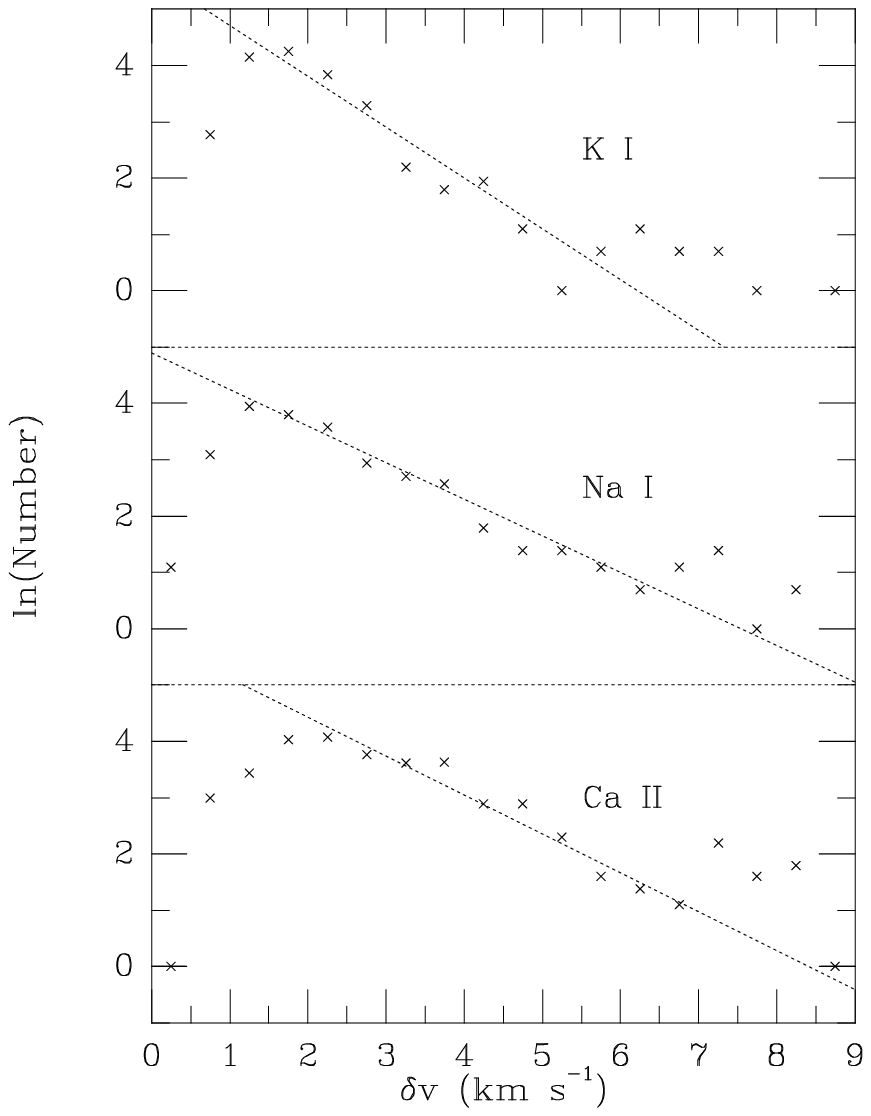}
\caption{The distribution of separations velocity separations ($\delta$v)
between adjacent absorption components for \NaI, \KI, and \CaII.  
The dotted line shows the linear fit 
over the range 2 \kms$\leq  \delta v \leq 6.5 $ \kms, with slope 
--0.9,  --0.69, for \KI, and \CaII\ respectively.
\protect\cite[Figures from][]{WeltyNa:1994,WeltyCa:1996,WeltyHobbsK:2001}
\label{fig-poisson}}
\end{figure}

The saturated \NaI\ D lines and the crowding of components in velocity space 
inspired observations of the weak
\NaI\ UV 3302 \AA\ doublet which fall on the linear
portion of the COG.
The mean widths for the 3302 \AA\ and strong D-line doublets
are $<$\bdoppler$>$=0.9 \kms\ and 1.5 \kms\ \cite[][]{Crutcher:1975}, indicating
that the weaker features reveal cloud cores, while the D-lines trace more extended
regions.
After correction for ionization effects, sodium was found to be depleted by a 
factor of $\sim$4--5.
During the last decade of the 20th century, a series of
high-resolution ($\sim$0.3--1 \kms) surveys of optical absorption lines
have provided a detailed look at the crowding of cloud components in velocity space.  
A high-resolution study of the \NaI\ D1 line towards 38 stars
found 276 absorption components \cite[][WHK]{WeltyNa:1994}. 
WHK developed a method of fitting 
complex absorption lines with individual cloud components characterized 
by component column density ($N$), velocity (\V), and velocity
dispersion (\bdoppler$^2$=2k\Tkinetic/$m$+2\Vturb$^2$).
The \NaI\ components have median values 
\bdoppler=0.73 \kms\ and \logNNaI=11.09 \cmtwo, and a
median separation between adjacent components of $\sim$2.0 \kms.
Weaker \NaI\ components exhibit a broader velocity distribution
than stronger components, displaced somewhat to negative velocities.
The correlation between \NHI\ and \NNaI\ decreases in quality for \NNaI$<$10$^{11}$ \cmtwo.
For internal cloud temperatures of 80 K, $>$38\% components have subsonic
internal turbulence for assumed Gaussian profiles for individual clouds.
WHK showed that component separations for \NaI\ are consistent
with a Poisson distribution (Figure \ref{fig-poisson}),
and that for this distribution only $\sim$60\% of the true total number of cloud 
components are being detected (given detection limits).
Similar results were found later for the distribution of \CaII\ and \KI\ components.

The alkali element Li is observed optically. 
A survey of the \LiI\ weak 6708 \AA\ (IP=5.4 eV)
towards 22 stars found seven positive detections at the 0.3--3 \mAA\ level
\cite[e.g.][]{WhiteLi:1986}.
The depletions of Li, C, Na, and K are correlated, and log $\delta \sim$--0.25 dex
in cold clouds (Figure \ref{fig-depl}).
Recently, the relative isotope ratio
$^7$\LiI/$^6$\LiI\ 
has been observed as a test of light element formation.
The $^7$\LiI/$^6$\LiI\ ratio appears to be variable, and for
at least one sightline ($o$ Per) cosmic ray spallation appears significant.

In cold clouds \CaII~is heavily depleted onto grains, and \CaII~is the dominant ionization state, while in warm gas Ca is only moderately depleted but \CaIII\ is the 
dominant ionization state (see Figure \ref{fig-zetaOph}).
The ratio \NCaII/\NCaI\ in individual velocity components yields 
\nel=0.055--0.57 \cc, independent of abundance uncertainties \cite[e.g.][]{WhiteCaI:1973}.  
If photoionization of trace elements is the sole source of electrons,
densities \nH$\sim$500--5,000 \cc\ are implied.  However, 
H may be partially ionized. 
Intercloud medium sightlines also show reduced line strengths of \NaI\ 
(factor of $>$5) in comparison to sightlines containing clouds.
High-spectral resolution (0.3--1.2 \kms) observations of the \CaII\ K line
provide a component-to-component sample of the relative properties of the \NaI\ and
\CaII\ profiles 
\cite[][WMH]{WeltyCa:1996}.
WMH found similar mean LSR velocities for
\CaII\ and \NaI\ (--1.7 \kms, --0.7 \kms), but larger velocity dispersions
for \CaII\ (12.3 \kms\ vs. 8.6 \kms).  The median \CaII\ column density
is 5 $\times$ 10$^{10}$ \cmtwo, and the median line width
is \bdoppler=1.31 \kms\ corresponding to \temp$\leq$4,100 K.
Based on line widths, less than 35\% of the \CaII\ components originate in warm gas
(6,000--8,000 K).  If \temp$>$500 K, most components have subsonic internal
cloud turbulent velocities.  Inside of the same cloud, \CaII\ is distributed over a
broader spatial region than \NaI, as shown by \bdoppler\ for \CaII\ and
\NaI\ (median values 0.84 and 0.65 \kms\ respectively).
The ratio \NNaI/\NCaII\ correlates with \NaI\ (with an offset at larger
column densities where \CaII\ is the dominate ionization state of Ca). 
Exceptions are sightlines showing the Routly-Spitzer effect, e.g. 
$\alpha$ Oph.  
Along low-density sightlines, up to 10\% of the total mass in gaseous
\CaII\ is at \V$>$10 \kms.

The component structure of \KI\ may provide an important clue to tiny cold clouds
since potassium is not subject to large depletion variations \cite{Hobbs:1974}.
Hobbs found that \NNaI, \NKI, \ebv, and \NHI\ (from \cop\ \Lalpha\ data)
are mutually correlated to within a factor of two, but
\NCaII\ did not correlate with any of these other quantities.
\NaI\ and \KI\ show a quadratic dependence on \NH:
(\NNaI$\sim$\NH$^{2.3}$, \NKI$\sim$\NH$^{2.0}$) indicating
that \nel/\nH\ is constant in the cloud and N$\sim$\nH$^2$ for photoionization.
The differential distribution of \KI\ cloud column densities ($\phi$($N$)) follows
a power law, $\phi$(\NKI)$\sim$\NKI$^{-3/2}$, giving
$\phi$(\NH)$\sim$\NH$^{-2}$.
About 4.6 \KI\ clouds kpc$^{-1}$ are seen.
A recent high-resolution survey (0.4--1.9 \kms) of \KI, with
2$\sigma$ detection limits of \NKI$\sim$1.4 $\times$ 10$^9$ \cmtwo,
provides the most complete available sample of cold ($\sim$80 K) diffuse
clouds. Observations of 54 stars yielded 319 components \cite{WeltyHobbsK:2001},
with the mean component velocity \Vlsr=1.7 \kms\ and
dispersion $<\V^2>^{1/2}$=7.5 \kms.  
Comparisons between \KI\ and other species revealed important trends.
\NNaI\ and \NKI\ show a linear relationship, with
similar median \bdoppler values (0.67 \kms\ and 0.73 \kms\ respectively). 
However, \CaII\ is formed over a larger velocity interval than \NaI\ or \KI, since
both the dispersion in bulk velocities, and line broadening, are larger for \CaII\ components than for
\NaI\ and \KI\ components.
The linear relationship between \NKI\ and \NCHI\ and similar profiles suggest 
these species are closely associated in dense clouds. 
\NKI\ and \NHtot\ are well correlated.
Median and maximum values for \NKI\ are 4 $\times$ 10$^{10}$ \cmtwo\ and 
10$^{12}$ \cmtwo, respectively, which convert to column densities 
\NH=2 $\times$ 10$^{20}$ \cmtwo\ and 10$^{21}$ \cmtwo, respectively, the observed \NKI--\NH\ correlation.
For \temp$\sim$80 K, 35--50\% of the components are found to have subsonic 
turbulence.
The \KI\ components are strongly crowded in velocity since
$\sim$85\% percent of the component separations (between two adjacent components)
are $<$3 \kms\
\cite[][and Figure \ref{fig-poisson}]{WeltyHobbsK:2001}.
Few sightlines are dominated by a single cloud.

The ion \TiII\ is unique among optically visible species because it is the 
dominant ionization state for Ti in diffuse clouds (IP$\sim$13.6 eV).
However, Ti has highly variable depletions in the ISM (Figure \ref{fig-zetaOph}).
The \TiII\ weak 3384 \AA\ line was surveyed towards 68 stars by \cite{Stokes:1978}. 
Titanium depletions vary by two orders of magnitude in the solar
neighborhood, with mean value \depl$_{\rm Ti}$$\sim$--2 dex.  
Comparisons between \NTiII\ and other species show that Ti and Ca
correlate more tightly with each other than other species 
(e.g. \NaI, \KI, \NH).  

The abundance properties of halo ISM were characteristized by a series of
studies of \CaII\ and \NaI\ towards
faint blue Feige stars near the Galactic poles \cite{CohenMeloy:1975}.
Large \RCaNa\ ratios in halo gas indicate 
intercloud type gas.  Halo gas velocities are 
skewed to the negative, and the infall requires the replenishment of halo gas on timescales $\sim10^8$ years.
Observations of halo stars found that \TiII\ is pervasive in the Galactic halo,
and Ti abundances increase with increasing \absZ\ \cite{Albert:1983}.  Albert concluded
that two types of gas contributed to the \TiII\ features: 
a thick low-velocity disk component (\absZ$<$200 pc), and 
a second high-velocity weakly depleted gas at high \absZ\
and constituting $\sim$24\% of the halo mass.
Low velocity gas shows Ti abundances $\sim$3\% solar, 
while for \absVlsr$>$10 \kms\ abundances average $\sim$6\% solar.
An extreme value is seen towards HD 123884, at \Z=8.7 kpc, where
the Ti abundance is 75\% solar.
The relations between $N{\rm sin}~|b|$ and \absZ~for \TiII, \CaII, \FeII, \HI, and \ebv\
show that the \TiII\ and \CaII\ gas is smoothly distributed and
extends to greater heights than other components,
with scale heights $>$2 kpc, 1 kpc, 0.5 kpc, 0.3 kpc, and  0.1 kpc, respectively,
for an exponential distribution \cite[Table \ref{tab-scaleH},][]{EdgarSavage:1989}.

Time-variable interstellar \CaII\ and \NaI\ absorption lines have been detected towards sources
in the Vela region \cite[e.g.][]{Hobbsetal:1991}.  These variable lines are characterized by
extremely large widths ($>$100 \kms) and ratios \NCaII/\NNaI\ (1.4 to over 56).
Variations are seen on timescales of 
$\sim$7 years towards HD 72127A, located near a 
bright filament in the Vela supernova remnant.

The velocity of the shocked interstellar gas towards $\alpha$ Oph, compared
to the velocity of ISM inside of the solar system led to the conjecture that
Loop I drives a flow of interstellar material past the Sun \cite{Frisch:1981}.
A series of papers modeling the kinematics of nearby ISM with a linear flow
find different flow vectors, depending on the star sample 
\cite[e.g.][]{Crutcher:1982,LallementVidalMadjarFerlet:1986,Vallergaetal:1993,Frisch:1995}.
Crawford found that 79\% of the components towards a set of
stars in Sco-Cen have negative LSR velocities consistent with a shell
expanding at $\sim$9 \kms\ around the
Upper Centaurus-Lupus subgroup of the Sco-Cen Association \cite{Crawford:1991}.
Although the nearest optical absorption lines show velocities generally
consistent with the cloud velocity inferred from observations of
\Lalpha\ and \HeII\ 584 \AA\ backscattered radiation, 
nearby stars often show more than one absorption component 
\cite[e.g., $\alpha$ Aql at 5 pc,][]{Ferlet:1999} and there is a distribution
of component velocities about the central flow velocity.
The heliocentric velocity vector for the cloud surrounding the solar system
(--25.3 \kms\ from the direction \glong=4\deeg, \glat=+16\deeg, )
corresponds to an LSR vector of --15 \kms\, approaching from \glong$\sim$346\deeg, \glat$\sim$--1\deeg\ (using Hipparcos solar apex motion).
Typical line broadening for \CaII\ K lines towards nearby stars
(\bdoppler$\sim$2.2 \kms)  are 
consistent with \Tkinetic$\sim$7,000 K and \Vturb$\sim$1 \kms\ 
\cite{CrawfordLallementWelsh:1998}.  However, several nearby stars
($\alpha$ Pav, $\delta$ Vel, $\delta$ Cyg) have narrow lines, 
\bdoppler$\sim$1 \kms\ indicating cold clouds ($<$600 K) within
24--50 pc of the Sun.

\subsection{Warm Ionized Material \label{WIM}}

Classic \HII~regions surrounding young massive stars in stellar associations
delineate the spiral arm structure of the Milky Way Galaxy 
\cite{GeorgelinGeorgelin:1976}.   
However, pulsar data and diffuse optical recombination lines 
show that diffuse warm low density interstellar plasma is widespread.

A comparisons of pulsar dispersion and rotation measures, for pulsars
with an independent distance estimate, showed that the
layer of free electrons in our Galaxy has a mean density
$<$\nel$>$$\sim$0.025 \cc\ and scale height $\sim$0.9 kpc \cite[e.g.][]{FalgaroneLequeux:1973}.
Taylor and Cordes (1993) produced a model for the distribution
of free electrons in our Galaxy which allowed for a non-uniform
distribution of plasma imposed by spiral arm structure and other 
large scale variations in the ISM. 
Several regions of enhanced diffuse \nel\ are seen, including towards
the Gum Nebula (240\deeg$\rightarrow$270\deeg) and the Sagittarius arm
(330\deeg$\rightarrow$30\deeg).
This model provides a means to estimate pulsar distances 
from dispersion measures \cite{TaylorCordes:1993}.

The mapping of diffuse (\nel$\sim$0.1 \cc) interstellar \Halpha~in our Galaxy is largely the result of work by Ron 
Reynolds and collaborators, who use large throughput high spectral resolution
Fabry-Periot techniques to show that faint \Halpha\ and 
and forbidden optical emission lines closely
follow the velocity distribution of \HI\ 21 cm emission visible in spiral
arms \cite[][]{Reynolds:1995}. 
The discovery of pulsars in globular clusters allowed the
comparison of pulsar dispersion measures with emission measures.
Reynolds obtained a midplane density $<$\nel$>$$\sim$0.08 \cc,
for \temp=7,000 K plasma which fills $\approx$20\% of a 2 kpc thick 
layer in the Galactic plane (known as the `Reynolds Layer').
Most of the $\Halpha$ luminosity in the ISM arises in traditional \HII\ regions
surrounding hot stars, however most of the interstellar \HII\ mass
($\sim$90\%) is associated with a diffuse extensive warm ($\sim$8,000 K),
low density ($\sim$ 0.1 \cc) component.
Diffuse \HII\ differs from classic \HII\ regions 
by large [\SiII]$\lambda$6716/\Halpha\ and small 
[\OIII]$\lambda$5007/\Halpha\ ratios.
The line widths of the metastable lines suggest $T\sim$8,000 K, while
the absence of \NI~indicates the underlying material is ionized (\nel/\nH$>$0.75).  
The spectrum of the ionizing source is constrained to be softer than
an O-star spectrum by
the failure to detect \HeI~$\lambda$5876 recombination, which indicates that
He is mainly neutral (\nHeII/\nHI$<$0.27) in the ionized WIM.

The local Galactic magnetic field has been mapped 
using the Faraday rotation of pulsar signals,
since \rotmeas/\dispmeas\ provides \Bparallel\
weighted by the electron density \cite{RandKulkarni:1989}.
Data from $\sim$200 pulsars show that RM's which sample
the North Polar Spur region are dominated by this feature. 
When the North Polar Spur contributions are removed from the sample,
a local magnetic field of \Bparallel$\sim$1.6$\pm$0.3 $\mu$G, directed towards 
\glong=96\deeg$\pm$4\deeg, is obtained.  
The Galactic field reverses towards the inner Galaxy at \dist=600$\pm$80 pc.
Rand and Kulkarni detected a contribution from 
random magnetic fields, $B \sim$5 $\mu$G, and cell length 55 pc, which 
they attributed to fields in unresolved shell structures.

Ionized rims are found on over 10\% 
of interstellar clouds \cite{Reynoldsetal:1995}.  
The distribution of WIM which
produces diffuse \Halpha\ emission correlates strongly with \HI\ 21 cm clouds:
$\sim$30\% of the diffuse \Halpha\ and $\sim$10--30\% of 21 cm emission 
occur in clouds with similar spatial distribution and kinematics, although
\Halpha\ and 21 cm emission are spatially separated as expected for cloud rims.
The clouds with rims typically are low density (\nH=0.2--0.3 \cc),
and have \NHI=2--20x10$^{\rm 19}$ \cmtwo\ and \EM=2--10 \cmsixpc.
The scale heights for clouds with ionized rims 
(\Z=0.1--1 kpc) exceeds the scale heights of cold \HI\ ($\sim$100 pc, 
Table \ref{tab-scaleH}).

\section{The Galactic Halo \label{iuehst}}

\subsection{Highly Ionized Gas \label{HIM}}

The most significant contribution of \iue\ to ISM studies
was the discovery and mapping of highly ionized gas towards faint stars in the Galactic halo and beyond.
The \CIV\ 1550 \AA\ and \SiIV\ 1402 \AA\ resonance
doublets were detected towards 
UV bright O and Wolf-Rayet stars (V=9--12 mag) in the 
Large and Small Magellanic Clouds \cite{SavagedeBoer:1981},
and $\sim$24
faint high latitude (\absZ$>$1 kpc) luminous OB stars in the Galactic halo 
\cite{PettiniWest:1982}.  
Hot halo gas extends 10--15 kpc below the Galactic plane
towards the LMC and SMC, and to velocities $\sim$150 \kms\ and  $\sim$50 \kms\
in the respective directions. 
Halo gas appears to share the rotation pattern of the underlying disk gas.
The interstellar origin of high-\Z\ hot gas was firmly established
by a survey of
over 40 faint (\dist$>$2 kpc) disk and halo B stars which were
selected to have no obvious association with circumstellar nebulosity
\cite{SavageMassa:1987}.  Scale heights, \scaleH, for the highly ionized
species were found to be
significantly larger than for neutral and weakly 
ionized species (Table \ref{tab-scaleH}, although recent evidence
for a patchy halo confuses the question).
The unambiguous detection of 
diffuse halo \NV\ was first made towards halo stars in the inner Galaxy
(\absZ=0.7--2.2 kpc, $R_{\rm g}<$6 kpc).  Savage and Massa found line strengths 
of \NV, \CIV, \SiIV\
an order of magnitude larger than expected for a photoionized halo
\cite[e.g.][]{SavageMassa:1987}.  

A series of \hst\ studies of halo gas, by 
Blair Savage, Ken Sembach, and collaborators,
showed that two types of highly ionized halo gas are present
\cite[e.g.][]{SavageSembach:1996}.
In Type I hot gas, \NSiIV:\NCIV:\NNV$\sim$1.0:3.0:0.5 and 
the narrow \SiIV\ and \CIV\ components are kinematically 
associated with low-ionization species, possibly originating in 
interface regions between warm and hot gas.
\CIV\ is found together with both the sharp \SiIV\ and broad \NV\ lines.
In Type II gas, minimal \SiIV\ is seen, with few low ions, and \NCIV/\NNV$\sim$1--3.
The broad \NV\ features are produced by a widely distributed component,
possibly from cooling hot gas in supernova bubbles or Galactic fountains.

A classic paper by Savage et al. (1997) combined observations of extragalactic
objects (QSOs and Seyfert galaxies) and 15 halo stars to probe the full
extent of HIM in the halo.
The derived exponential scale heights showed that high ions 
extend a factor of $\sim$10 further from the plane than \HI\
(Table \ref{tab-scaleH}). 
Ionization process must vary with \absZ, since the 
ratios \RCIVNV~and \RSiIVNV~are enhanced at mid-\absZ\
distances by a factor of 2, compared to lower and higher \absZ\ values.
\CIV~and \SiIV~correlate well, and both correlate poorly with \NV.
Halo HIM corotates with disk gas out to \absZ$\sim$5 kpc for $R_{\rm galactic}$$>$5 kpc. 
HIM halo gas shows typical net inflow velocities of
$\sim$20 \kms, which is about twice that of neutral gas.
The turbulent broadening of \CIV\ ($\sim$60 \kms) is inadequate
by factors of $\sim$3 to support halo gas, indicating that additional
pressure sources such as the 
Galactic magnetic field or cosmic rays are required.

Observations of \AlIII, \SiIV, \CIV, and \NV\ 
showed that profiles of \AlIII, a tracer of photoionized gas, are 
generally narrower than those of \SiIV, \CIV, and \NV\ in halo gas
\cite[e.g.,][]{SembachSavage:1992}.
The `apparent optical depth technique' was used to derive information on 
unresolved components or saturation in the \iue\ profiles.  
Highly ionized gas extends to greater distances from the Galactic
plane than \AlIII, which has a scale height comparable to the
Reynolds layer thickness ($\sim$1 kpc).
\NV\ absorption was detected in 80\% of the sightlines,
indicating pervasive \T$\sim$2 $\times$ 10$^5$ K collisionally ionized gas in the
galactic halo.
These relatively low resolution \iue\ data (25 \kms) show relatively
constant mean ratios of \RCIVSiIV=3.6$\pm$1.3 and \RCIVNV=4.6$\pm$2.7.

Neither \iue\ nor \hst\ could observe \OVI.
The far UV spectrometer on an ORFEUS-SPA mission observed \OVI\ towards
Galactic and extragalactic sightlines 
and found \OVI\ to be significantly more confined to the plane than 
\SiIV, \CIV, or \NV\ \cite{HurwitzBowyer:1996}.
The North Polar Spur region (Loop I and Loop IV) are sources of enhanced
coronal gas, supplying hot gas to the Galactic halo.
Variations in the \OVI/\NV\ ratio reveals the non-uniform distribution of 
high latitude coronal gas.
A small \OVI\ excess was tentatively identified towards two stars located behind 
an intermediate negative velocity cloud (\absVlsr=45--70 \kms)
identified by Danly (1992).
Relative line strengths are consistent
with either collisionally ionized gas in a Galactic fountain, or composite
mixture of conductive interfaces and turbulent mixing layers.
Recently, FUSE has confirmed many of the overall characteristics of halo \OVI\ found by ORFEUS-SPA.
FUSE observations of \OVI~towards 11 AGN's confirm that
\OVI\ is more concentrated towards the Galactic plane than
\SiIV, \CIV, or \NV, and that the scale height of high ions
decreases with increasing ionization  \cite[][e.g. see Table \ref{tab-scaleH}]{Savageetal:2000}.
$N$(\CIV)/$N$(\OVI) ranges from $\sim$0.15 in the disk to $\sim$0.6 in the halo.
Variations in the \Z-component of \NOVI\ (factor $\sim$2.5)
confirm that \OVI\ halo gas is highly patchy.
\hst\ observations of \CIV\ towards $\mu$ Col combined with \cop\
observations of  \OVI\ show these features occur at the same velocity, but
\OVI\ is twice as broad as \CIV\ \cite{Brandtetal:1999}. 
The ratio \NCIV/\NOVI=0.11$\pm$0.1, and is typical of hot disk material.

Production models for highly ionized halo gas are still uncertain \cite{Spitzer:1996}. 
Possible origins include conductively heated interfaces, radiative cooling,
or turbulent mixing layers.  The conductive heating models
require contact between cooler material and
hot plasma (10$^6$ K gas), such as found inside supernova remnants
or stellar wind bubbles.
A representative radiative cooling model would be a Galactic fountain, 
with the infall of cooling plasma onto the Galactic plane.
The turbulent mixing layer model mixes
hot and cold gas in a turbulent shear flow of hot gas past cold clouds.
Conductive heating models reproduce \NCIV/\NOVI$\sim$0.15 found at
low \absZ, but are unsuccessful at high $z$.  
Smaller scale heights for more highly ionized gas suggest an origin
for \OVI\ connected to disk supernova remnants \cite{Savage:1995}.

\subsection{Abundances in Halo Gas \label{haloISM}}

When Supernova 1987A (SN1987A) exploded in the LMC, it was quickly observed
as a target-of-opportunity by \iue\ and provided a brief brilliant
opportunity to observe halo gas using a 
$V\sim$4.5 mag background source with a featureless continuum \cite[e.g.][]{Bladesetal:1988,WeltySN:1999}.
Intermediate velocity halo gas (56$\leq$\Vlsr$\leq$90 \kms) 
was seen to be a mixture of warm and cool gas,
with virtually no depletion of the refractory elements Si, Cr, Mn, Fe, Ni and Fe
relative to Zn.  The high velocity halo gas (109$\leq$\Vlsr$\leq$140 \kms)
showed typical abundance characteristic of warm partially ionized
halo gas ($T\geq$4500 K).

A series of \hst\ studies of the ISM towards halo stars by Spitzer, Savage,
Fitzpatrick, Sembach, and others show that depletion patterns
in halo and disk are similar, except that the least
depleted sightlines generally are towards halo stars. 
Zn depletions in individual clouds towards halo stars range from 
$\delta_{\rm Zn} \sim -0.4 \rightarrow0.0$ dex for \nHmean$<$0.1 \cc\ for \absVlsr$<$50 \kms\ \cite[][]{Fitzpatrick:1996}.
Both warm disk and halo clouds typically show solar abundances of S 
and at least one IVC (the --70 \kms\ component towards HD 93521)
shows supersolar abundances for S
\cite[perhaps from partial H ionization][]{SavageSembach:1996,Fitzpatrick:1996,HowkSavageFabian:1999}.
Typically $>$10--30 cloud components are seen towards halo stars.
In low velocity clouds ($<$20--25 \kms) there is a clear positive 
correlation between gas phase abundances and increasing cloud velocity for 
refractories (Si, Fe, Cr, Mn, Ti).  The strongest variations are found for Fe ($\sim$1.5 dex)
and the smallest variations for Si (0.5 dex).
Below \absVlsr$\sim$20 \kms, the gas-phase abundances of refractory elements
(Fe, Mn, Ni, Si, and Cr) generally increase with velocity, while
above that value depletions tend not to vary with velocity.
Fe and Si depletions correlate strongly for both disk and halo stars,
with maximum abundances in disk and halo clouds
corresponding to  $\delta_{\rm Fe} \sim$--0.5 dex, and $\delta_{\rm Fe} \sim$--0.15 dex \cite[with respect to S, ][]{Fitzpatrick:1996,SavageSembach:1996}.
For \nHmean$>$0.1 \cc, Cr/Zn$\sim$--1.4 dex and is density independent for
disk stars.  
For \nHmean$<$0.1 \cc, Cr depletions are in the range
log $\delta_{\rm Cr} \sim$--1.6 to --0.8 dex (with respect to H).
Warm halo gas components towards HD 116852 (\dist=4.8 pc, \Z=--1.3 kpc)
show that gas-phase abundances increase with \absZ,
and line broadening (velocity dispersion) and ion scale heights
increase as the ionization of the gas increases \cite{SembachSavage:1996}.

\begin{figure}[h]
\vspace*{4in}
\includegraphics{PCFfig6.eps}
\caption{Depletions in disk and halo gas 
'H', 'W', 'C' indicate Halo gas, Warm disk
clouds, and Cold disk clouds based on \hst\ data.  Depletions from
identified clouds towards $\zeta$ Oph, 23 Ori, $\zeta$ Ori, and $\mu$ Col are labeled.
WLV, SLV, HVC refer to weak low velocity, strong low velocity and high velocity
clouds towards 23 Ori \protect\cite{Welty23:1999}.
(Figure courtesy of Dan Welty.) \label{fig-deplHWC}}
\end{figure}

\input{PCFtabscalehts}

\subsection{Warm Neutral and Ionized Gas \label{WM}}

Comparisons between \iue\ observations of low and high ions and
\HI\ 21 cm data, show halo kinematics, and constrain cloud distances
\cite[e.g.,][]{Danly:1992}.
The low halo (\absZ$<$1 kpc) contains approximately equal numbers
of negative and positive velocity components, while the distant halo
has a preponderance of infalling gas.
In the Northern Hemisphere, low ion (e.g. \SiII) lines have component
velocities of up to \absVlsr$<$120 \kms.  Intermediate
velocity gas ($>$40 \kms) is formed at \absZ$>$1 kpc, and
has an excess of negative velocity components.  
Low velocity gas ($<$40 \kms), with
an equal number of positive and negative components,
tends to be closer to the Galactic plane (\absZ$<$1).  
\hst\ observations of halo stars, which typically sample long sightlines devoid of dense cold clouds, provide some of the best data on warm gas.
\hst\ found that up to 30 or more absorption components may be seen towards
halo stars with  \Z$\sim$2 kpc, extending over large velocity intervals ($>$100 \kms).
Such sightlines sample both disk and halo gas.

Towards HD 100340, nine regions of intermediate velocity ($+31\rightarrow+78$ \kms)
partially ionized gas are seen, with column densities in the range 
\logNHI=18.38--19.36  \cmtwo.
\HI\ 21 cm data show these clouds are
warm neutral gas, $T\sim$7,000 K (from the FWHM values).
Typical densities are \nHI=0.022--0.10 \cc.
Towards HD 93521 four low velocity components (\Vlsr$<$20 \kms) show
\nel$\sim$0.11 \kms, while intermediate velocity gas (\Vlsr$>$25 \kms)
have \nel$\sim$0.04 \kms, based on \CIIstar/\CII\ \cite[see][for references]{Fitzpatrick:1996}.
The 30 components seen towards HD 215733 (\Z=1.7 kpc)
show temperatures $\le$300 K$\rightarrow$$>$1000 K and electron densities \nel=0.02 to 0.06 \cc.

In individual warm disk and halo clouds, high and low ionization levels are commonly associated with each other.
For example, towards HD 215733
\SiIV\ and \CIV\ lines are at the velocities of 
three warm clouds, although line widths indicate hot
temperatures (\temp=6 $\times$ 10$^5$ K and 5 $\times$ 10$^4$ K, respectively).
The neutral weakly ionized IV (+70 \kms) cloud towards HD 20366 
is warm (5,300-6,100 K)
with \nel=0.15-0.34 \cc, however highly ionized gas
with \NCIV/\NSiIV$\sim$4.5 is present suggesting collisionally ionized cloud 
interfaces with \T$\sim$10$^5$ K \cite[e.g.,][]{Sembach:1995}.

Observations in a number of warm clouds towards halo stars 
(e.g. HD 215733, HD 154368) show that 
different diagnostics for \nel\ in the same cloud yield
discrepant results (as also seen for cold clouds, Section \ref{diskISM}).
The ionization equilibrium of \CI, \MgI, \SI, and \CaII\ give 
log \nel\ differing by up to one dex, with generally \nel(\MgI)$>$\nel(\rm \CI),
when the temperature dependence of \MgII\ and \CII\ recombination rates
are assumed similar.
These differences are consistent with the trace neutral status of \CI\ and \MgI\
in cool gas, and enhanced \MgI\ in warm gas ($>$5,000 K) from dielectronic recombination.
In addition, \CaII\ and \TiII\ (Section \ref{opticalhires}) are distributed relatively smoothly in the halo, and \TiII\ extends to \Z$>>$2 kpc.
Electron densities based on \NCaII\ vary smoothly
in comparison to \nel\ derived from either \NNaI\ or \NMgI/\NMgII, which
generally sample cloudy regions \cite{CardelliSembachSavage:1995}.

Systematic variations in cloud electron densities as a function
of velocity have been detected towards stars in the low and high halo.
The stars $\mu$ Col and HD 93521, for example, show electron densities 
which decrease with increasing cloud velocity
\cite[e.g.,][]{SpitzerFitzpatrick:1993}.
Component \nel\ values found from both \CIIstar/\CII\ and \CaII\ show a
clear decrease with increasing \absVlsr towards HD 93521.
A spread in \nel\ values of 0.02$\rightarrow$0.05 \cc\ is found
for \absVlsr$\sim 50 \rightarrow $70 \kms.
Five warm neutral velocity components are seen towards $\mu$ Col \cite[--29 $\rightarrow$ +41 \kms][]{ HowkSavageFabian:1999}.
Warm intermediate velocity gas (+31,+41 \kms, $T$=4000$\pm$700 K) is
low column density (\logNH=17.3--17.8 \cmtwo).
Electron densities in component 4, for example, 
are \nel=0.47$\pm$0.14 \cc\ (using \NCIIstar/\NCII)
and \nel=0.64$\pm$0.14 \cc\ (using \CII\ and \NMgI/\NMgII).

UV \Lalpha\ data provide the distribution of \HI, but generally do not
distinguish between cold and warm gas (in contrast to \HI\ 21 cm
data where both absorption and emission data are available).
\HI\ scale heights were determined from IUE observations of 393 stars
($d$=0.12--11 kpc, \absZ$<$4 kpc) which are
free of stellar \Lalpha~contamination \cite{DiplasSavage:1994}.  
Selecting sightlines which
avoided obvious clouds yielded scale height \scaleH=195 pc, and an overall 
value \nHmean=0.23 atoms \cc\ (range 0.017--8.62 atoms \cc).
Color excess data indicate that dust is patchier than the gas,
with dust scale height \scaleH=152 pc and midplane value \ebv/\dist=0.257 mag kpc$
^{\rm -1}$.
Assuming, rather, a two-component exponential model to accommodate
gas clumping around the plane gives
for the compact component \nH$_{\rm ,1}$(0)=0.247 \cc, \scaleH$_1$=73 \pc,
and for the extended component \nH$_{\rm ,2}$(0)=0.16 \cc, \scaleH$_2$=357 \pc.
These scale heights are comparable to 21 cm results
despite different underlying assumptions and sightline lengths.
Denser clouds show a larger ratio of \NHI/\ebv\ than towards interarm
sightlines, possibly indicating grain modification in dense clouds.
Savage (1995) finds that $\sim$82\% of the halo
interstellar mass is contributed by neutral gas (both the confined and extended
components).  An extended component of warm ionized gas
provides $\sim$15\% of  the mass, while the extended hot component 
contributes $ \sim$3\% of the halo ISM mass.

\section{Closing Comments}

What does the 21st century hold for ISM studies?
Opening up the UV window on our Galaxy initiated the golden age of space astronomy,
and revealed 30--40\% of the Galactic mass previously invisible.
Will the retirement of \hst\ shutter this window?
The 21st century may see the first {\it in situ} measurements of the 
interstellar cloud surrounding the solar system \cite{ISP:2000}.
When we leave our heliosphere and explore interstellar space we will
become citizens of the Milky Way Galaxy.  However many problems are
solvable with earth-orbiting spacecraft.  The next instrumental advance for UV astronomy 
will be to launch an ultrahigh resolution UV space spectrometer 
(912--3000 \AA) with a large dynamic range (V=1--12 magnitudes) to reveal the physics of cloudlets in the disk and halo. 

What are the questions of the future?

\begin{enumerate}

\item Where will the solar journey take us, and what
are the past and future Galactic environments of the Sun? 

\item Is the ISM one grand, turbulent, gravitationally-layered continuum, 
with locally variable
properties sculpted by spiral density waves which initiated epochs of star-formation?

\item Will better data on Galactic halo gas unlock the mysteries of chemical evolution in our Galaxy?

\item Is the ISM composition homogeneous?  Is Seaton's postulate that 
dust grains contain the atoms missing from the ISM gas phase correct?

\item Where does halo ISM originate?

\item Are molecular clouds an ISM reservoir replenishing diffuse gas?
  
\item What is the answer to Parker's question: ``What is an
interstellar cloud''? 
Are turbulence, cloud interface regions, and expanding
superbubble shells relevant for the answer to this question?

\end{enumerate}

Finally, we return to Harlow Shapley's idea.  Are encounters with
interstellar clouds important for evaluating the historical climate
of the Earth, and by extrapolation climate stability for extra-solar planets?

$Acknowledgements$
The author would like to thank Dan Welty for providing many of the figures
in this paper, and both Dan Welty and Don York 
for numerous informative scientific discussions. This research 
has been supported by NASA grants NAG5-6405 and NAG5-7077.
This paper is appearing in $The~Century~of~Space~Science$,  
Eds. J. Bleeker, J. Geiss, and M. C. E. Huber, Kluwer Academic Publishers, 2001.

\newpage

\end{document}

%% file: PCFdefcommand.tex
\def\SrII{\hbox{\ion{Sr}{2}}}
\def\BaII{\hbox{\ion{Ba}{2}}}

\def\ArI{\hbox{\ion{Ar}{1}}}

\def\AlI{\hbox{\ion{Al}{1}}}

\def\AlIII{\hbox{\ion{Al}{3}}}
\def\CHI{\hbox{\ion{CH}{1}}}
\def\CHII{\hbox{\ion{CH}{2}}}
\def\NCHI{\hbox{$N$(\ion{CH}{1})}}

\def\CI{\hbox{\ion{C}{1}}}
\def\CIstar{\hbox{\ion{C}{1}$^{*}$}}
\def\CIstarstar{\hbox{\ion{C}{1}$^{**}$}}
\def\CII{\hbox{\ion{C}{2}}}
\def\CIIstar{\hbox{\ion{C}{2}$^*$}}

\def\CIV{\hbox{\ion{C}{4}}}
\def\ClI{\hbox{\ion{Cl}{1}}}
\def\ClII{\hbox{\ion{Cl}{2}}}
\def\NClI{\hbox{$N$(\ion{Cl}{1})}}
\def\NClII{\hbox{$N$(\ion{Cl}{2})}}
\def\SIV{\hbox{\ion{S}{4}}}
\def\CNI{\hbox{\ion{CN}{1}}}
\def\CaI{\hbox{\ion{Ca}{1}}}
\def\CaII{\hbox{\ion{Ca}{2}}}
\def\CaIII{\hbox{\ion{Ca}{3}}}
\def\LiI{\hbox{\ion{Li}{1}}}

\def\CrII{\hbox{\ion{Cr}{2}}}

\def\FeI{\hbox{\ion{Fe}{1}}}
\def\FeII{\hbox{\ion{Fe}{2}}}

\def\HH{\hbox{H$_2$}}
\def\HD{\hbox{HD}}
\def\HI{\hbox{\ion{H}{1}}}
\def\DI{\hbox{\ion{D}{1}}}
\def\HII{\hbox{\ion{H}{2}}}
\def\HeII{\hbox{\ion{He}{2}}}

\def\HeI{\hbox{\ion{He}{1}}}
\def\KI{\hbox{\ion{K}{1}}}
\def\KII{\hbox{\ion{K}{2}}}

\def\PII{\hbox{\ion{P}{2}}}
\def\MgI{\hbox{\ion{Mg}{1}}}
\def\MgII{\hbox{\ion{Mg}{2}}}
\def\MnII{\hbox{\ion{Mn}{2}}}
\def\NI{\hbox{\ion{N}{1}}}
\def\NII{\hbox{\ion{N}{2}}}

\def\NV{\hbox{\ion{N}{5}}}
\def\Na{\hbox{Na}} 
\def\RNaCa{\hbox{\mbox{$N$(\ion{Na}{1})/$N$(\ion{Ca}{2})}}} 
\def\RCaNa{\hbox{\mbox{$N$(\ion{Ca}{2})/$N$(\ion{Na}{1})}}}

\def\RCIVSiIV{\hbox{\mbox{$N$(\ion{C}{4})/$N$(\ion{Si}{4})}}} 
\def\RCIVNV{\hbox{\mbox{$N$(\ion{C}{4})/$N$(\ion{N}{5})}}} 
 
\def\RSiIVNV{\hbox{\mbox{$N$(\ion{Si}{4})/$N$(\ion{N}{5})}}} 
\def\NaI{\hbox{\ion{Na}{1}}} 
\def\NaII{\hbox{\ion{Na}{2}}} 
\def\NiII{\hbox{\ion{Ni}{2}}}
\def\CuII{\hbox{\ion{Cu}{2}}}
\def\OI{\hbox{\ion{O}{1}}}

\def\OIII{\hbox{\ion{O}{3}}}

\def\OVI{\hbox{\ion{O}{6}}}
\def\SII{\hbox{\ion{S}{2}}}
\def\SI{\hbox{\ion{S}{1}}}
\def\SIII{\hbox{\ion{S}{3}}}
\def\SiII{\hbox{\ion{Si}{2}}}

\def\SiIII{\hbox{\ion{Si}{3}}}
\def\NSiIII{\hbox{$N$(\ion{Si}{3})}}
\def\NSiII{\hbox{$N$(\ion{Si}{2})}}
\def\SiIV{\hbox{\ion{Si}{4}}}

\def\TiII{\hbox{\ion{Ti}{2}}}

\def\ZnII{\hbox{\ion{Zn}{2}}}

\def\iue{\hbox{IUE}}
\def\ghrs{\hbox{GHRS}}
\def\cop{\hbox{$Copernicus$}}
\def\hst{\hbox{HST}}
\def\vsini{\hbox{$vsini$}}
\def\scaleH{\hbox{$h$}}
\def\dist{\hbox{$d$}}

\def\bdoppler{\hbox{$b_{\rm Dop}$}}

\def\Msun{\hbox{$M_{\rm sun }$}}
\def\2S12{\hbox{$^{2}{\rm S }_{1/2}$}}

\def\EM{\hbox{EM}}
\def\DM{\hbox{DM}}

\def\emmeas{\hbox{$\int$  $n_{\rm e}^2$ d$L$}}
\def\dispmeas{\hbox{$\int$ $n_{\rm e}   dL$}}
\def\rotmeas{\hbox{$\int$$n_{\rm e}$ $B_{\rm \parallel} dL$}}
\def\Z{$z$}

\def\absZ{\hbox{$|z|$}}

\def\V{\hbox{$V$}}
\def\Vturb{\hbox{$v_{\rm turb}$}}

\def\Vlsr{\hbox{$v_{\rm lsr}$}}
\def\Vhc{\hbox{$v_{\rm hc}$}}

\def\Vradial{\hbox{$v_{\rm radial}$}}
\def\absVlsr{\hbox{$\mid{}v_{\rm lsr}\mid{}$}}

\def\ebv{\hbox{$E(B-V)$}}
\def\W{\hbox{$W$}}

\def\Bparallel{\hbox{$B_{\parallel}$}}

\def\B{\hbox{$B$}}
\def\Halpha{\hbox{H${\alpha}$}}

\def\Lalpha{\hbox{Ly${\alpha}$}}
\def\Lbeta{\hbox{Ly${\beta}$}}

\def\Ldelta{\hbox{Ly${\delta}$}}
\def\Lepsilon{\hbox{Ly${\epsilon}$}}

\def\cmsixpc{\hbox{cm$^{-6}$ pc}}

\def\au{\hbox{{\rm AU}}}

\def\temp{\hbox{$T$}}
\def\T{\hbox{$T$}}
\def\Tbright{\hbox{$T_{\rm bright}$}}
\def\Tkinetic{\hbox{$T_{\rm kinetic}$}}

\def\Tex{\hbox{$T_{\rm excitation}$}}
\def\Tspin{\hbox{$T_{\rm spin}$}}
\def\Tcond{\hbox{$T_{\rm cond }$}}

\def\muG{\hbox{$ \mu{\rm G}$}}
\def\pc{\hbox{pc}}
\def\NX {\hbox{$N$(X)}}

\def\NSII{\hbox{$N$(\ion{S}{2})}}

\def\NSiII{\hbox{$N$(\ion{Si}{2})}}
\def\NSiIII{\hbox{$N$(\ion{Si}{3})}}

\def\NFeII{\hbox{$N$(\ion{Fe}{2})}}
\def\NCaI{\hbox{$N$(\ion{Ca}{1})}}
\def\NCaII{\hbox{$N$(\ion{Ca}{2})}}
\def\nNaI{\hbox{$n$(\ion{Na}{1})}}
\def\nNaII{\hbox{$n$(\ion{Na}{2})}}
\def\NNaI{\hbox{$N$(\ion{Na}{1})}}
\def\NKI{\hbox{$N$(\ion{K}{1})}}
\def\NNaII{\hbox{$N$(\ion{Na}{2})}}
\def\logNNaI{\hbox{log$N$(\ion{Na}{1})}}
\def\NOVI{\hbox{$N$(\ion{O}{6})}}

\def\NCIV {\hbox{$N$(\ion{C}{4})}}
\def\NNV {\hbox{$N$(\ion{N}{5})}}
\def\NSiIV {\hbox{$N$(\ion{Si}{4})}}
\def\NSIV {\hbox{$N$(\ion{S}{4})}}

\def\kms{\hbox{km s$^{\rm -1}$}}

\def\deplX{\hbox{$\delta_{\rm X}$}}
\def\depl{\hbox{$\delta$}}
\def\glong{\hbox{$l^{\rm II}$}}
\def\glat{\hbox{$b^{\rm II}$}}

\def\absglat{\hbox{$\mid{}b\mid{}$}}
\def\logNHI{\hbox{log$N({\rm H^o})$}}

\def\logNH{\hbox{log$N({\rm H})$}}

\def\fHH{\hbox{$f_{\rm H_2}$}}

\def\NH{\hbox{$N$(H)}}
\def\NDI{\hbox{$N$(\ion{D}{1}$)$}}
\def\NHI{\hbox{$N$(\ion{H}{1}$)$}}
\def\NHII{\hbox{$N($\ion{H}{2}$)$}}
\def\NHH{\hbox{$N$(H$_{\rm 2}$)}}

\def\NOI{\hbox{$N({\rm O^o})$}}
\def\NNI{\hbox{$N({\rm N^0})$}}

\def\NCI{\hbox{$N({\rm C^o})$}}

\def\NCII{\hbox{$N({\rm C^+})$}}
\def\NCIIstar{\hbox{$N({\rm C^{+ *}})$}}
\def\NMgI{\hbox{$N({\rm Mg^{o}})$}}
\def\NMgII{\hbox{$N({\rm Mg^{+}})$}}
\def\NTiII{\hbox{$N({\rm Ti^{+}})$}}

\def\NHtot{\hbox{$N$(${\rm H^o }$)+2$N$(${\rm H_2}$)}}

\def\nHH{\hbox{$n({\rm H_2 }$)}}
\def\nHI{\hbox{$n({\rm H^o})$}}
\def\nH{\hbox{$n_{\rm H}$}}

\def\nHmean{\hbox{$\overline{n_{\rm H}}$}}
\def\nHImean{\hbox{$\overline{n{\rm (H^o) }}$}}
\def\nHHmean{\hbox{$\overline{n_{ \rm H_2}}$}}
\def\nelmean{\hbox{$\overline{n_{\rm e}}$}}

\def\nHII{\hbox{$n({\rm H^+})$}}

\def\nHeII{\hbox{$n({\rm He^+})$}}
\def\nel{\hbox{$n_{\rm e}$}}

\def\N{\hbox{$N$}}

\def\mAA{\hbox{m\AA}}
\def\cmtwo{\hbox{cm$^{-2}$}}

\def\cc{\hbox{cm$^{-3}$}}

\def\deeg{\hbox{$^{\rm o}$}}

%% file: PCFtabscalehts.tex
\begin{table}[ht]
\caption{Stratification of Interstellar Gas away from the Galactic Plane \label{tab-scaleH}}
\begin{center}
\begin{tabular}{llccll}
\multicolumn{2}{c}{Component$^{\rm 1}$}&Mid-Plane& \scaleH$^{\rm 2}$ & Method$^{\rm 3}$&Reference\\
&&Density&  & &\\
&&(\cc)& (kpc) & &\\

\hline
\HH&&0.14&0.07&UV& Savage et al. (1977) \\
\HI&CNM&0.70&0.07 & UV& Bohlin et al. (1978) \\
   &WNM&0.16&0.36 & UV& Diplas and Savage (1994) \\
\HI&CNM&0.39&0.11g&21 cm&Dickey and Lockman (1990) \\
   &WNM&0.11&0.25g&21 cm& Dickey and Lockman (1990) \\
   &WNM&0.06&0.40 &21 cm& Dickey and Lockman (1990)\\
\HII&WIM &0.014& 0.07& DM& Reynolds (1995) \\
    &WIM &0.024& 0.9& DM& Reynolds (1995) \\
\FeII&&&0.5 &UV& Edgar and Savage (1989)\\
\AlIII&WIM&&1.0$^{+0.36}_{-0.24}$  &UV& Savage et al. (1990)\\
\CaII$^{\rm 4}$&&&1 & Opt.& Edgar and Savage (1989) \\
\TiII$^{\rm 4}$&&&$>$2 &  Opt.& Edgar and Savage (1989) \\
\OVI&HIM &2.0 $\times$ 10$^{\rm -8}$& 2.7$\pm$0.4 & UV& Savage et al. (2000)\\
\NV&HIM  &2.0 $\times$ 10$^{\rm -9}$& 3.9$\pm$1.4 & UV& Savage et al. (1997)\\
\CIV&HIM &9.2 $\times$ 10$^{\rm -9}$& 4.4$\pm$0.6 & UV& Savage et al. (1997)\\
\SiIV&HIM&2.3 $\times$ 10$^{\rm -9}$& 5.1$\pm$0.7 & UV& Savage et al (1997) \\
\hline
\end{tabular}
\end{center}
Notes:
1.  CNM, WNM, WIM, and HIM are cold neutral, warm neutral, warm ionized,
and hot ionized material, respectively.  
2.  Based on Savage (1995). 
The scale height \scaleH\ is defined by the density distribution
as a function of distance to the galactic plane, \Z, and generally
is described by an exponential distribution:
$n$(\Z)=$n$(0)exp(--\Z/\scaleH) \cc, where $n$(0) is the midplane density
and \scaleH\ is the scale height.   Distributions labeled by `g' are
Gaussian: $n$(\Z)=$n$(0)exp(--\Z/\scaleH)$^2$/2. 
3.  Methods are:  UV = ultraviolet absorption lines; 21 cm = \HI\ hyperfine
line; DM = pulsar dispersion measures; Opt. = optical absorption lines.
4.  Enhanced abundances of \CaII\ and \TiII\ in the WNM yield smoother
distributions than for the CNM. 
\end{table}